\documentclass{article}

\usepackage{arxiv}
\usepackage{dirtytalk}
\usepackage[utf8]{inputenc} 
\usepackage[T1]{fontenc}    
\usepackage{hyperref}       
\usepackage{url}            
\usepackage{booktabs}       
\usepackage{amsfonts}       
\usepackage{nicefrac}       
\usepackage{lipsum}
\usepackage{chngcntr}\usepackage{graphicx}
\usepackage{dcolumn}
\usepackage{stackengine}
\usepackage[section]{placeins}
\usepackage[dvipsnames*,svgnames]{xcolor}
\usepackage{wrapfig}
\definecolor{AGblue}{rgb}{0.0, 0.375, 0.75}  
\newcommand{\beginsupplement}{%
        \setcounter{table}{0}
        \setcounter{section}{0}
        \setcounter{subsection}{0}
        \renewcommand{\thetable}{\arabic{table}}%
        \setcounter{figure}{0}
        \renewcommand{\thefigure}{S\arabic{figure}}%
     }     
     
\usepackage{verbatim}
\usepackage[position=bottom,caption=false,captionskip=3pt,font=normalsize,subrefformat=simple,labelformat=simple]{subfig}

\usepackage{listings}
\usepackage{mathtools}
\usepackage{todonotes}
\usepackage{amsmath}
\usepackage{chngcntr}
\usepackage{bm}
\usepackage{algorithm}
\usepackage{algpseudocode}
\usepackage{authblk}
\usepackage[super]{nth}
\usepackage[euler]{textgreek}
\renewcommand{\mathbf}{\boldsymbol}
\usepackage{caption}

\hypersetup{
    colorlinks=true,
    linkcolor={blue},
    filecolor={blue},      
    urlcolor={blue},
    citecolor={blue}
}

\usepackage[outercaption]{sidecap}

\definecolor{AGblue}{rgb}{0.0, 0.375, 0.75}  
\newcommand{\AG}[1][]{\textcolor{AGblue}}
\newcommand{\red}[1][]{\textcolor{red}}
\newcommand{\new}[1][]{\textcolor{blue}}

\title{Super-Suppression of Long Phonon Mean-Free-Paths in Nano-Engineered Si due to Heat Current Anticorrelations}

\author[1,$\dagger$]{S. Aria Hosseini}
\author[2,$\dagger$]{Alathea Davies}
\author[1]{Ian Dickey}
\author[3,*]{Neophytos Neophytou}
\author[1,*]{P. Alex Greaney}
\author[2,*]{Laura de Sousa Oliveira}

\affil[1]{Department of Mechanical Engineering, University of California, Riverside, Riverside, CA 92521, USA}
\affil[2]{Department of Chemistry, University of Wyoming, Laramie, WY 82071, USA }
\affil[3]{School of Engineering, University of Warwick, Coventry, CV4 7AL, United Kingdom}
\affil[$\dagger$]{These authors contributed equally to this work}
\affil[*]{Correspondence: NN: N.Neophytou@warwick.ac.uk; PAG: greaney@ucr.edu; LdSO: Laura.deSousaOliveira@uwyo.edu}

\begin{document}
\maketitle

\begin{abstract}
The ability to minimize the thermal conductivity of dielectrics with minimal structural intervention that could affect electrical properties is an important capability for engineering thermoelectric efficiency in low-cost materials such as Si. We recently reported the discovery of special arrangements for nanoscale pores in Si that produce a particularly large reduction in thermal conductivity accompanied by strongly anticorrelated heat current fluctuations~\cite{PhysRevB.102.205405} -- a phenomenon that is {missed by the diffuse adiabatic boundary conditions conventionally used in Boltzmann transport models.} This manuscript presents the results of molecular dynamics simulations and a Monte Carlo ray tracing model that teases apart this phenomenon to reveal that special pore layouts elastically backscatter long-wavelength heat-carrying phonons. This means that heat carriage by a phonon before scattering is undone by the scattered phonon, resulting in an effective mean-free-path that is significantly shorter than the geometric line-of-sight to the pores. This effect is particularly noticeable for the long-wavelength, long mean-free-path phonons whose transport is impeded drastically more than is expected purely from the usual considerations of scattering defined by the distance between defects. This \say{super-suppression} of the mean-free-path below the characteristic length scale of the nanostructuring offers a route for minimizing thermal conductivity with minimal structural impact, while the stronger impact on long wavelengths offers possibilities for the design of band-pass phonon filtering. Moreover, the ray tracing model developed in this paper shows that different forms of correlated scattering imprint a unique signature in the heat current autocorrelation function that could be used as a diagnostic in other nanostructured systems.
\end{abstract}

\section{Introduction}

Advancements in nanoengineering provide unprecedented control over phonon-mediated heat transport and enable obtaining low thermal conductivity in materials, such as Si (an inexpensive, abundant and non-toxic semiconductor), that are intrinsically high thermal conductors. This ability is of special importance for realizing low-cost thermoelectrics for thermal harvesting and energy conversion applications. An extreme reduction in the thermal conductivity through nanostructuring has been observed in several Si-based membranes, e.g., nanowires~\cite{ma2019quantifying, li2012thermal, doi:10.1021/acs.nanolett.6b02450}, thin films~\cite{cheaito2012experimental,braun2016size}, nanocomposites~\cite{https://doi.org/10.48550/arxiv.2110.13375,miura2015crystalline,liao2015nanocomposites}, superlattices~\cite{hu2012si,mu2015ultra,garg2013minimum}, nanoporous alloys~\cite{liu2020thermoelectric,shi2018polycrystalline,PhysRevB.100.035409,lim2016simultaneous,https://doi.org/10.48550/arxiv.2203.13279} and amorphous structures~\cite{doi:10.1063/5.0054039}. Interfaces in the nanoscale structures scatter phonons truncating the distance over which phonons convey heat --- the phonon mean free path (MFP). However, reductions in thermal conductivity can come at a penalty to electrical transport properties~\cite{doi:10.1021/acsaem.0c02640}, and so it is important to find strategies for reducing the MFP of phonons with minimal structural intervention, especially for applications such as thermoelectrics.

{Thermal transport in nanoporous materials has been heavily researched. It has been shown that the room temperature thermal conductivity of Si-based nanoporous materials can be reduced beyond a material’s amorphous limit~\cite{doi:10.1021/acsnano.5b05385, doi:10.1021/nl102931z}. Surface area~\cite{PhysRevB.98.115435, doi:10.1063/1.2817739, doi:10.1021/nn2003184}, number of (pore) scatterers~\cite{PhysRevB.91.054305}, size and shape of pores~\cite{doi:10.1021/nl802045f, LIU20101547}, pore spacing and distribution~\cite{WEI2020104619, lee2017investigation, doi:10.1021/acsami.8b00097, PhysRevB.96.115425}, boundary roughness~\cite{doi:10.1063/1.4879242}, and amorphicity~\cite{doi:10.1063/1.4948337} are many of the geometric features that have been investigated. In most geometries, strong phonon-boundary scattering and a reduction in the line-of-sight, i.e. a blocking in the phonon pathways along the transport direction, have largely explained the reduction in thermal conductivity in nanoporous materials. The main characteristic of these prior works is that scattering happens at the characteristic length scale of the introduced disorder, the mean free path is reduced accordingly, and the reduction in thermal conductivity follows Matthiessen's rule to a large degree~\cite{PhysRevB.98.115435}. As phonons with mean free paths on the disorder length are impeded more, it is the shorter wavelength (larger wavevector) phonons that are affected more, whereas the long wavelength acoustic phonons are affected the least~\cite{PhysRevB.98.115435, CHAKRABORTY2020109712}. The effect we describe is different. Physically it causes heat trapping between the pores which results in the annihilation of the backscattered heat flux. A super-suppression of the heat flux is observed as a result. Because the long-lived fluctuations correspond to low frequencies, the close-packed pores function as a thermal band-pass filter.}

In a recent study, we identify specific morphologies/arrangements of cylindrical pores in Si that produce a particularly large reduction in thermal conductivity, in some cases even pushing the conductivity below the amorphous limit~\cite{PhysRevB.102.205405}. Equilibrium molecular dynamics (MD) simulations unveiled a large thermal resistance via strong anticorrelation (AC) of the heat flux fluctuations, resulting in a suppression of the thermal conductivity to as much as 80\% lower than the thermal conductivity if the same pores are in a uniformly distributed arrangement. This is largely due to elastic backscattering of long-wavelength phonons by narrowly spaced pores, which was confirmed by a set of wavepacket collisions simulations that indicate that heat can oscillate back and forth between pores. In this manuscript, we present a theory that provides a deeper understanding of the scattering process that gives rise to anticorrelated fluctuations in the heat flux. We observe that even diffusive scattering can lead to an anticorrelated behavior with the same signature as that observed in nanoporous Si --- our initial guess was that only specular scattering is strong enough to produce this behavior. Perhaps most importantly, we demonstrate that the AC heat flux indicates that the truncated effective MFP, the median MFP of the thermal conductivity distribution~\cite{hosseini2021nondiffusive}, is smaller than the average line-of-sight, {the distance a phonon can travel without being scattered~\cite{doi:10.1063/1.4993601}}, between pores. We use the term \say{super-suppression} to refer to this truncation of the phonon MFP below the characteristic length scale of the nanostructuring. Super-suppression occurs through correlated scattering---{backscattering of phonons without thermalization}---such that reflected phonons undo some of the heat propagation by the phonons prior to scattering.  The selective scattering of long-wavelength phonons, which correspond to larger MFPs, implies that the close-packed pores act as a band-pass filter.

\begin{figure}[t]
    \centering
    \includegraphics[width=1.0\textwidth]{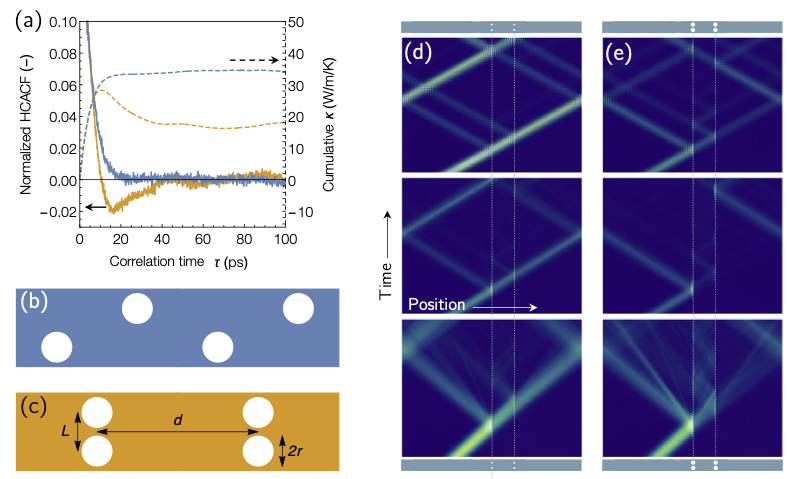}
    \caption{Plot (a) shows the HCACF (solid lines, left-hand axis) and HCACF integral (dashed lines, right-hand axis) with the color-coding corresponding to the two porous geometries’ arrangements (panels b\&c)  shown in the lower panel. Both geometries have identical porosity, but the pores are offset in the blue geometry and stacked in the orange geometry. Plots (d) and (e) show position-time kinetic energy maps for phonon wavepackets with various wave vectors colliding with two ranks of pores. The location of the pores is indicated with the vertical dashed white lines. The plots in column (d) are for pores with r = 3~\AA\ and (e) for pores with r = 7~\AA. In both cases, the top row is for long-wavelength phonons (near $\Gamma$), the middle row is for phonons with wave vectors about halfway between $\Gamma$ and the Brillouin zone edge, and the bottom row is for wave vectors about three-quarters of the way to the Brillouin zone edge. The larger pores show stronger backscattering, and that phonons scatter back and forth multiple times between the two ranks of pores.}
    \label{fig:fig1}
\end{figure}

The remainder of the manuscript is structured as follows. Section~\ref{sec:motivating-observations} describes the observations that motivate this study. Section~\ref{sec:Results-and-Discussion} presents our main results and discussion. We first discuss the results from the molecular dynamics calculations (Sec.~\ref{sec:Molecular-Dynamics-Model-HCACF}), followed by the derivation of the ray tracing Monte Carlo (MC) model and the discussion of the insights obtained from it (Sec.~\ref{sec:theory-bulk-conductivity}). The MD calculations include a set of equilibrium simulations to compute the thermal conductivities of Si with various arrangements of pores using the Green-Kubo method~\cite{PhysRevB.103.224204, PhysRevMaterials.6.015403} followed by the spectral analysis of the heat current (Sec.~\ref{sec:HCACF_Spectral_Analysis}). These calculations indicate that in geometries where the AC effect is present, long mean-free-path, and thus long-wavelength, phonons are selectively suppressed or filtered beyond typical scattering behavior leading to a non-monotonic heat current autocorrelation function (HCACF). We compared MD simulations for 40 distinct geometries using a statistical model based on Matthiessen's rule (Sec.~\ref{sec:Super-Suppression of Phonon MFP}). This comparison between close-packed and non-close-packed porous geometries provides further evidence that the correlated scattering that emerges in close-packed porous geometries leads to the super-suppression of the effective phonon MFP below the characteristic length-scale of the porous structure. To interpret the heat current autocorrelation function, an analytical model was developed that relates the thermal conductivity and heat current autocorrelation function to the occupancy autocorrelation function of individual phonon modes (Sec.~\ref{sec:Stochastic-Model-of-Heat-Current} \& Appx.~\ref{sec:apndx-theory-bulk-conductivity}). This model was used as the basis for a Monte Carlo ray tracing scheme to compute the heat current correlation function from a population of phonons experiencing different forms of correlated scattering (Sec.~\ref{sec:Monte-Carlo}). The ray tracing model shows that different forms of scattering imprint signatures in the HCACF that could be used as a diagnostic in other nanostructured systems. In quantifying the effective phonon MFPs in both the MD and ray tracing models, we show that correlated scattering leads to the MFP being super-suppressed, even if the correlation between incident and reflected phonons is lost after more than one scattering event. We also provide the analysis of the cross-correlation function of the spatially decomposed heat flux to further illustrate how heat is trapped between rows of pores. This analysis reveals that the behavior observed in the MC ray tracing model is consistent with diffuse scattering (Appx.~\ref{sec:Heat-Flux-Decomposition}).
\section{Motivating Observations --- Anticorrelation and Anomalously Low Thermal Conductivity} \label{sec:motivating-observations}

Figure~\ref{fig:fig1} shows the heat current autocorrelation and resulting cumulative thermal conductivity for two different pore configurations; one with pores aligned adjacently to form a picket (in gold), the other staggered (in blue). The calculations were performed using equilibrium MD simulations, the details of which are described in Supplementary Information (SI) Sec.~\ref{sec:computational-methods}. The pores in the two materials are identical in shape and size, thus the materials have the same porosity and number density of pores. However, when the pores are aligned the thermal conductivity is approximately one-half of the one in which pores are staggered. In the aligned configuration, heat has to pass through a narrow constriction, and this will increase thermal resistance. However, the magnitude of resistance observed in the MD simulations is much larger than the resistance one would predict from a simple geometric argument. {This effect is not observed in calculations that numerically solve the spatially- and mode-dependent Boltzmann transport equation (BTE)~\cite{romano2021openbte}, due to the use of diffuse adiabatic boundaries in conventional Boltzmann transport models.} Contrary to the trend observed using molecular dynamics, calculations of the thermal conductivity of Si containing the two different pore layouts in Fig.~\ref{fig:fig1} using BTE solver OpenBTE~\cite{romano2021openbte} predicts the thermal conductivity of the material with aligned pores to be larger than that with the staggered pores by a factor of $\mathrm{\sim}$1.16. The anomalously large thermal resistance of the aligned pores in the MD simulations is accompanied by a qualitatively different trend in the HCACF which has a region of negative, i.e., anticorrelated, correlation. This implies that heat current fluctuations traveling in one direction are frequently followed by fluctuations traveling in the opposite direction. We show below that the AC behavior arises from correlated reflections of phonons at the surface of the pores --- backscattering without thermalization --- which means that the heat carried by phonons before the collision with the pores is partially undone by the backscattered phonons. The BTE calculations described above fail to predict the enhanced thermal resistance of the aligned pores because, although they capture ballistic transport effects across the full phonon spectrum in Si and explicitly model the pore geometries, they do not include the backscattering phenomenon. The pore surfaces in these calculations were treated as diffusely scattering adiabatic boundaries with a model that assumes that phonons thermalize with the phonon bath when they are scattered.

{Anticorrelated heat current fluctuations have been reported in a variety of materials due to several different underlying mechanisms. In metal-organic frameworks (MOFs), the flapping modes due to linkers oscillating back and forth cause negatively correlated heat flux fluctuations~\cite{ZHOU2021100516, PhysRevMaterials.6.015403}. The negative regimes of the HCACF in MOFs are reflected in the convective heat flux term. The same can be said for liquids, where convective atomic motion (i.e., mass transport) results in negative HCACF minima~\cite{PhysRevE.62.2188}. In crystal structures, the convective heat flux is negligible and only the virial term contributes to the thermal conductivity. Layered materials with significant atomic mass differences exhibit similar behavior because the center of kinetic energy oscillates but the center of mass doesn't; this is seen, for instance, in HfB\textsubscript{2}~\cite{doi:10.1063/1.3647754}. The anticorrelated HCACF regions are also observed in amorphous materials~\cite{PhysRevB.103.224204, MCGAUGHEY20041783}. McGaughey and Kaviany attribute this behavior to distinct local environments. These mechanisms are distinct from the diffusive backscattering of the heat flux in the crystalline close-packed nanoporous geometries discussed in this work. Moreover, in these examples, the HCACF dip is short-lived and located in the sub-picosecond region, whereas the HCACF dip in the crystalline porous geometries we study here is long-lived (tens of picoseconds). Probably anticorrelated heat current fluctuations in carbon nanotubes (CNTs) identified in Haskins et al.~\cite{haskins2014equilibrium} is the only study akin to ours. In their work, the dip location and depth change as a function of nanotubes length. The authors attributed this behavior to the elastic scattering reflection of the heat flux at the boundary. While the physics of the behavior they observed in the CNTs are comparable to the porous Si, the geometries (one-dimension with boundary reflections versus three-dimensions and backscattering from internal pores) and their applicability are vastly different.}

The anticorrelation effect seen in our MD simulations can be very large as seen in our previous work~\cite{PhysRevB.102.205405} where reported pore arrangements resulted in a final thermal conductivity, $\mathrm{\kappa_{\infty}}$, as low as 20\% of the peak cumulative thermal conductivity, $\mathrm{\kappa_{\mathrm{peak}}}$. This implies that the motion of phonons after scattering can undo up to 80\% of the heat conduction by the phonons before they were scattered. The pore arrangement influences the anticorrelation behavior in two ways: the periodicity of the pores along the transport direction, $d$, controls the lifetime over which the phonons’ momentum is correlated, while the lateral spacing, $L$, and the pore radius, $r$, determine the width of the necks between pores which sets the backscattering probability. This is illustrated in Fig.~\ref{fig:fig1}(b). These plots show kinetic energy heat maps of phonon wavepackets in Si with two pores 50 nm apart. The radius/neck is small in the left-hand plot while the right-hand plot has a large radius/necking ratio. We infer that a large portion of the transmitted phonon at the first pore is backscattered from the second one and trapped between them. In the study of the AC effect presented below we have performed MD simulations of systems with neck sizes of up to 6 nm, which represent geometries that are small, but close to being experimentally realizable. However, MC ray tracing simulations predict that the AC effect should also be present in much larger geometries --- systems too large to simulate with MD, but that are readily experimentally accessible with current nanofabrication techniques. We thus think that the detailed understanding of the backscattering and MFP super-suppression that we study in the remainder of this manuscript can help guide materials engineers in the design and development of new thermoelectric materials for energy applications.

\section{Results and Discussion}\label{sec:Results-and-Discussion}

Our study focuses on two materials systems: silicon and a {pseudo-material} with a grey phonon population that we model in our ray tracing simulations. The latter system enables us to examine the effect of correlated scattering in isolation and unobfuscated by the presence of a broad phonon spectrum.

\subsection{Molecular Dynamics Model HCACF}\label{sec:Molecular-Dynamics-Model-HCACF}

{The Fourier transform of the HCACF gives us the proportion of heat transported by phonons with a given longevity, and so} we begin by analyzing the HCACF spectra for a host of geometries with and without heat current anticorrelations (ACs) to more clearly illustrate how the AC affects the phonon MFP ($\lambda$). {When 
AC effects are present, we observe a sharp decrease in the HCACF spectra for long-lived phonons. These phonons have long MFPs and their suppression leads to unusually low thermal conductivities. These phonons also mostly have longer wavelengths and thus their reduction indicates the occurrence of long-wavelength band-pass filtering.} Two {analytical} models based on Matthiessen’s rule are proposed for geometries exhibiting the AC effect and standard geometries (without AC) based on the presence or absence of backscattering. We obtain a good agreement between the model-predicted thermal conductivities and the thermal conductivities computed using the Green--Kubo approach for both cases. Finally, heat flux cross-correlations are computed for bulk and a close-packed porous geometry, completing the picture of what happens in these geometries as heat becomes trapped and showing that while phonons in Si are long-lived enough to undergo multiple scattering events, they lose (temporal) coherence, {the process when phonons preserve their phase after a scattering event~\cite{PhysRevB.90.014307}}, after a secondary scattering in the close-packed geometries. The details on the molecular dynamics calculations are included in SI Sec.~\ref{sec:computational-methods}.

\subsubsection{HCACF Spectral Analysis} \label{sec:HCACF_Spectral_Analysis}

\begin{figure*}[t]
    \centering
    \includegraphics[width=1\textwidth]{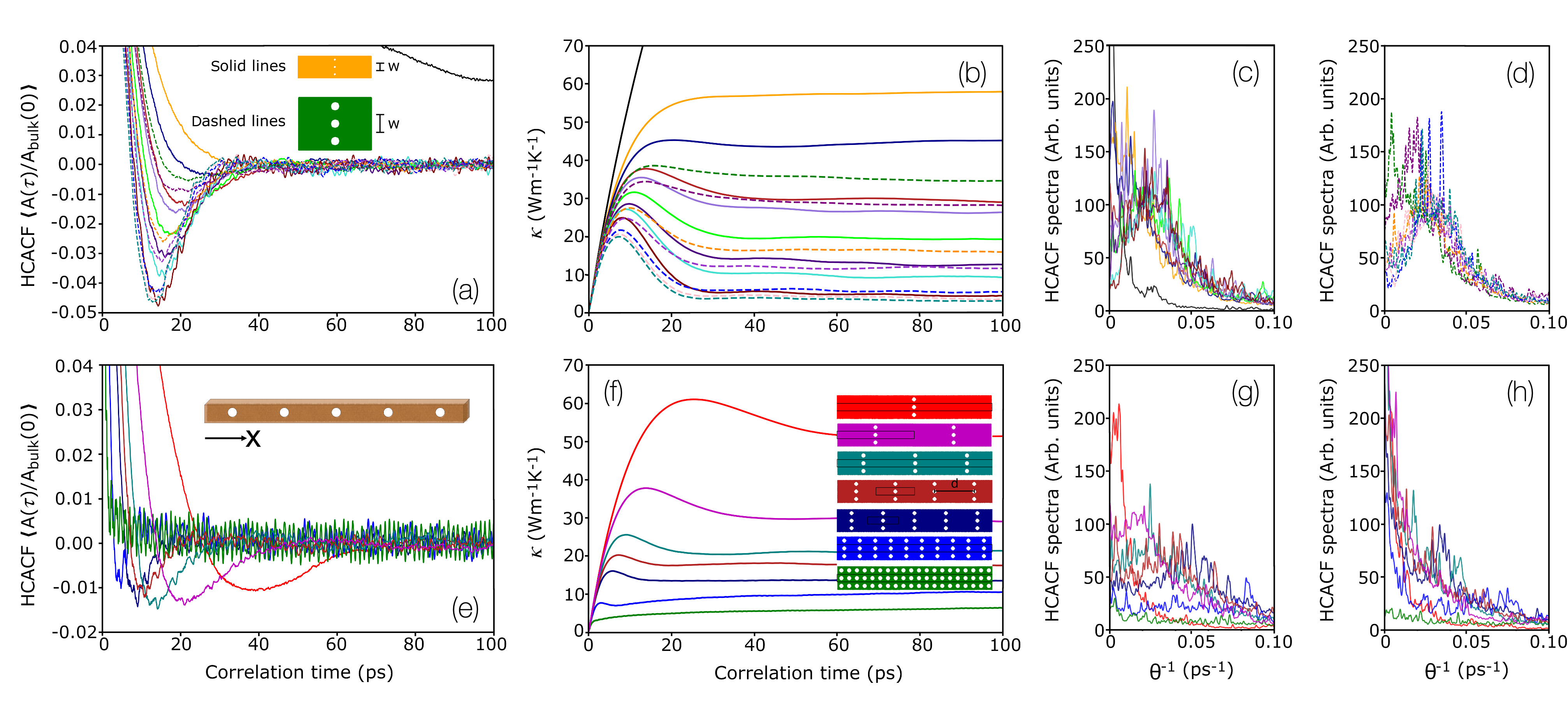}
    \caption{(a) {Moving average of the HCACF for two sets of $\mathrm{100\times10\times10}$ (solid lines) and $\mathrm{100\times24\times10}$ (dashed lines) supercells. (b) Cumulative thermal conductivity calculated using Eq.~\eqref{eq:kappa_GK}.} This shows the effect of the negative correlations that emerge in the HCACF on the thermal conductivity. Panels (a) and (b) are reproduced from Ref.~\cite{PhysRevB.102.205405}. (c) Low-frequency region of the HCACF power spectra, below 0.1 THz, for the $\mathrm{100\times10\times10}$ supercell set of geometries. (d) Same figure as (c) but for the $\mathrm{100\times24\times10}$ supercell set of geometries. (e) Moving average of the HCACF for the geometries shown as insets in pane (f). These geometries all have the same pore sizes, but the horizontal spacing between the pores varies. (f) Cumulative thermal conductivity calculated using Eq.~\eqref{eq:kappa_GK} for the same geometries. (g) Low-frequency region of the HCACF power spectra for the same geometries along the transport direction [indicated by an arrow in the inset in panel (e)]. (h) Low-frequency region of the HCACF power spectra for the same geometries perpendicular to the transport direction.}
    \label{fig:spectra}
\end{figure*}

Figures~\ref{fig:spectra}(a\&b) show the running-average of the normalized HCACF and the cumulative thermal conductivity as a function of correlation time, $\tau$, for two sets of geometries: a set with a narrow width --- 10 unit cells perpendicular to the transport direction --- marked with solid lines, and a set with a wider width --- 24 unit cells perpendicular to transport direction --- marked with dashed lines. The pore radii were varied from 1 nm up to 2.5 nm and from 3.6 nm up to 5.9 nm for the structures with narrow and wide widths, respectively. See Fig.~\ref{fig:SI-spectra}(a) in the SI for the complete set of simulations. The heat current anticorrelation effect is observed in structures with large pore radius to neck (packing) ratios. As the packing ratio increases, so does the amplitude of the dip in the HCACF [Fig.~\ref{fig:spectra}(a)]. A particularly large AC effect can be observed in Figs.~\ref{fig:spectra}(a\&b), where the HCACF dip magnitude reached a normalized value of $\sim$ 0.05, corresponding to a fraction of 0.2 of the thermal conductivity at the peak of the integrated HCACF [Fig.~\ref{fig:spectra}(b)].
 
{Figures.~\ref{fig:spectra}(c, d, g \& h) show the Fourier transform of the HCACF, i.e., the power spectrum of the heat current. The spectra tell us the distribution of lifetimes of heat current fluctuations, $\theta$.} For the configurations with no AC effect, the power spectrum decreases monotonically and proportionally to the length of the phonon MFP (note that $\lambda \propto \tau^{-1}$). However, as the dip in the autocorrelation appears, so does an inflection point in the spectra, below which the phonons' contribution to thermal transport is strongly suppressed and thus a non-monotonic behavior emerges. This is an evidence of the supper-suppression effect and indicates that phonons with long intrinsic MFPs (i.e., long fluctuation lifetimes) have been filtered out. The non-monotonic behavior with a peak in the HCACF spectra suggests that in structures with the AC, larger phonon trajectories [left-region of Figs.~\ref{fig:spectra}(c and d)] are back-scattered for a longer period of time and therefore more significantly suppressed, compared to shorter trajectories, which exhibit less back-scattering (approaching the peak height from the left).

In the presence of the supper-suppression, the intensity of the spectra in the large phonon MFP limit is a function of the HCACF dip amplitude. As the HCACF dip amplitude increases, the large MFP phonons' contribution to thermal transport before the \say{inflection point} decreases. This suggests that the AC effect arises from the scattering of the large-MFP acoustic modes at the surface of the pores. {These modes are known to have the highest contribution to thermal conductivity [see Fig.~\ref{fig:SI-spectra}(d)]. The truncation of the low-frequency end of the spectrum appears to be accompanied by a slight bulging in the region just above the inflection point. It is possible that the phonons in this range would otherwise have been scattered by the lower MFP phonons. Because the longer lived phonons have been cut off, due to the AC effect, fluctuations in this range contribute more significantly to the overall thermal conductivity.}

For the configurations with the AC effect, the location of the inflection points in the power spectrum is correlated with the HCACF dip minima. In Fig.~\ref{fig:spectra}(g), in which the spectra corresponds to geometries exhibiting HCACF dips at different correlation times (Fig.~\ref{fig:spectra}(f)), the maxima (or peaks) of the spectra shift to the left as the dips shift to the right. However, for the geometries in the panel (a), where the correlation times of the HCACF dips are roughly constant,  the spectral peaks show up at similar $\tau^{-1}$. The time scales corresponding to where the dips in the HCACFs and cumulative thermal conductivities occur are controlled by the spacing between the pores along the transport direction. We have previously reported that the correlation time at which the HCACF is most negative changes with the distance $d$ between the ranks of pores, and is consistently characterized by a single effective velocity $v_\mathrm{dip}$, which is found to be $\mathrm{\sim0.6}$ of the averaged acoustic phonon velocity~\cite{PhysRevB.102.205405}. This can be demonstrated by considering a new set of simulations with the same $\mathrm{100\times10\times10}$ supercell box size shown in Fig.~\ref{fig:spectra}(a) (yellow geometry) and a varying pore periodicity along the direction of transport, as shown in the inset of Fig.~\ref{fig:spectra}(f). The geometries inset in Fig.~\ref{fig:spectra}(f) match the HCACFs and cumulative thermal conductivities in Figs.~\ref{fig:spectra}(e) and ~\ref{fig:spectra}(f), respectively. In all cases, pores radius is 1.5 nm, and the neck is 2.43 nm. In Fig.~\ref{fig:spectra}(g), the {inflection point} shifts to a higher $\tau^{-1}$ (smaller MFPs) region as the spacing between the pores along the transport direction decreases. Figure~\ref{fig:spectra}(g) corresponds to the HCACF spectra in the direction of transport, while Fig.~\ref{fig:spectra}(h) corresponds to transport perpendicular to the length of the cells. Figure~\ref{fig:spectra}(g) shows both how the low thermal conductivity in these sets of simulations with packed pores is in part due to higher porosity (standard scattering behavior), but also due to a super-suppression effect, while Fig.~\ref{fig:spectra}(h) exhibits only standard scattering behavior. When we shrink the spacing between the pores, we see that the overall contribution of the low-frequency phonons to thermal transport decreases (as expected), corresponding to uniformly lowering the slope of the HCACF power spectrum, instead of an abrupt cutoff below an inflection point. No inversion of the slope is seen, unlike what happens in the longitudinal transport direction [Figs.~\ref{fig:spectra}(c\&d)] due to super-suppression. In Figs.~\ref{fig:spectra}(e\&f), of the small neck geometries, all but the green geometry exhibit a dip in the HCACF. By considering the HCACFs spectra (in $x$ and $y$), it can be seen why this is the case. The contribution of the large MFP modes has already been significantly reduced, and the necking AC effect is therefore eclipsed. Moreover, the spacing between the pores in the green geometry is identical in both directions (perpendicular to the pores). A broader region of the HCACF spectra and phonon lifetime versus frequency are shown in the SI figure~\ref{fig:SI-spectra}.

\subsubsection{Super-Suppression of Phonon MFP}\label{sec:Super-Suppression of Phonon MFP}

In the previous work~\cite{PhysRevB.102.205405}, we modeled the total thermal conductivity in Si with a uniform distribution of pores with a simple Matthiessen's rule expression
\begin{equation}
 \frac{1}{\kappa_p(\lambda)} = \frac{1}{\kappa_\mathrm{bulk}(\lambda)}\left(1 + \lambda\frac{\mathrm{C}}{d}\right).
\label{eqn:model}
\end{equation}
Here, $\kappa_{p}(\lambda)$ and $\kappa_{\mathrm{bulk}}(\lambda)$ are the materials' thermal conductivity due to the fraction of the phonon population with MFP of $\lambda$. In Eq.~\eqref{eqn:model}, $\lambda$ is scaled by the distance between the pores along the direction of transport, $d$, and a single numerical parameter $\mathrm{C}$ that is $\lambda$ independent. This simple formula could predict the total thermal conductivity within a mean absolute percentage error of 11.4\%. The term $d/\mathrm{C}=\Lambda_\mathrm{eff}$ is the effective $x$-projected distance that a phonon travels until it collides with a pore, and it depends on the phonon's view factor, {the possibility of a phonon traveling through the material without colliding with the pores~\cite{LIU20101547, hosseini2021enhanced}}, and the spacing between the pores. For a phonon heading along $x$ to a palisade of cylindrical pores, the probability that it passes between the pores unimpeded is $1-\alpha$, where $\alpha=2r/L$ is the pore fraction of the boundary, and $L$ is the width (along the pore stacking direction) of the supercell. The mean distance that a phonon travels along $x$ until it first encounters a row of pores is $(\nicefrac{1}{2})d$, so the distance traveled by phonons that pass through $n-1$ successive rows of pores before being stopped by the $n$\textsuperscript{th} pore is $\left(\frac{1}{2}+n\right)d$. Averaging over all $n$ weighted by the probability gives an effective mean free path of 
\begin{equation}
\label{eq:lambda_c_no_AC}
\Lambda_\mathrm{eff}=\sum_{n=1}^\infty d\left(\frac{1}{2}+n\right)\left(1-\alpha\right)^n\alpha = d \left(\frac{1}{\alpha}-\frac{1}{2}\right).
\end{equation}
The mean flight to encounter a pore is equivalent to the phonon effective MFP for pore scattering if the scattering at the pore is uncorrelated, in which case we would expect $\mathrm{C}=2\alpha/\left(2-\alpha\right)$. If on the other hand the phonon is back scattered such that the effective MFP for pore scattering is super-suppressed below the distance to the pores we could expect 
\begin{equation}
\label{eq:lambda_c_AC}
\Lambda_\mathrm{eff}=\sum_{n=1}^\infty d n\left(1-\alpha\right)^n\alpha = d \left(\frac{1}{\alpha}-1\right),
\end{equation}
which gives $\mathrm{C}=\alpha/\left(1-\alpha\right)$.

\begin{SCfigure}
\includegraphics[width=0.5\textwidth]{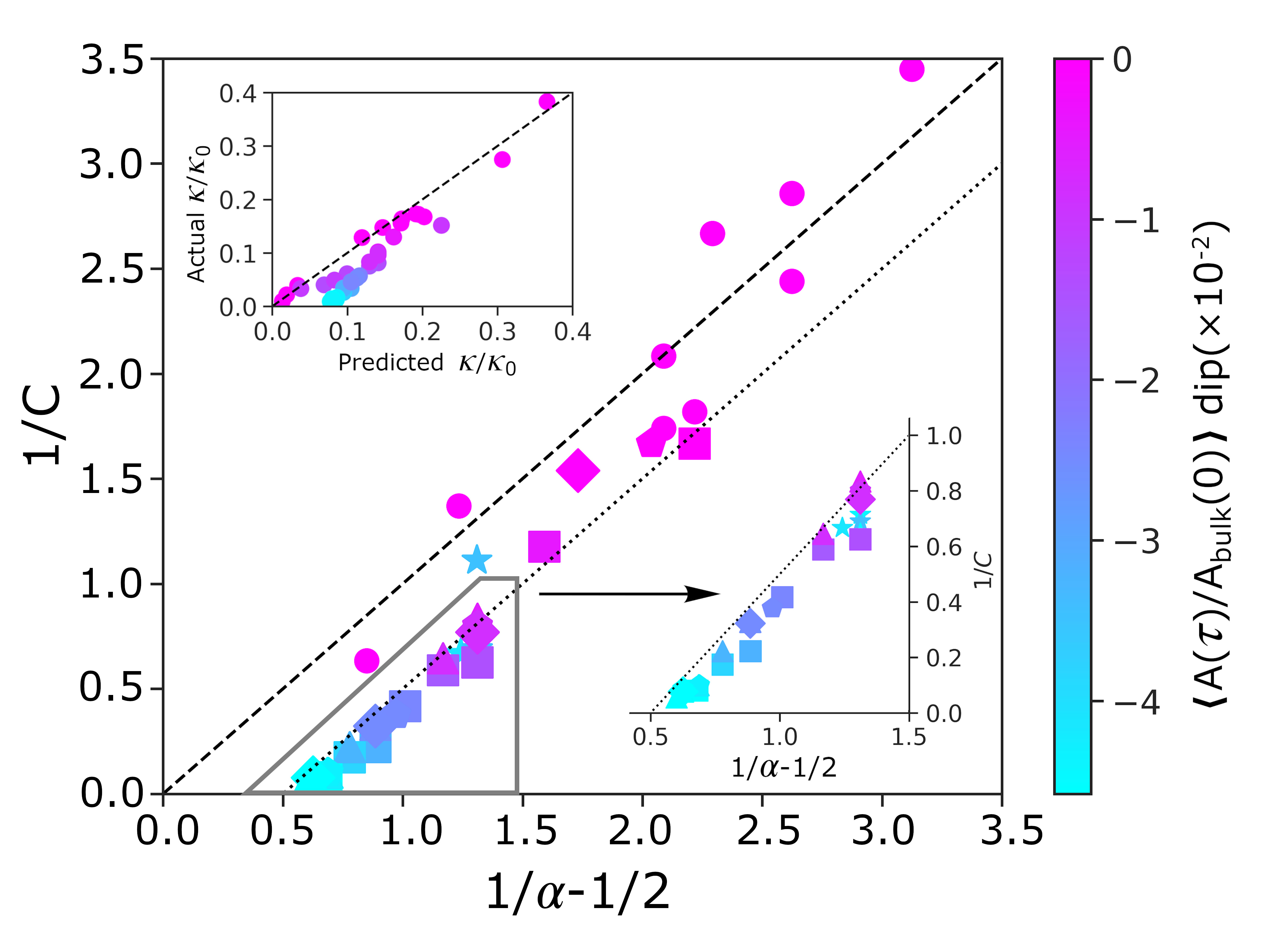}
\caption{The plot of the scattering strength constant, $\mathrm{C}$, as a function of the pore fraction for typical scattering conditions (dashed line), and super-suppression (dotted line). Each datum corresponds to a given geometry (with varying pore sizes and distributions) and the color scheme maps to the depth of the HCACF dip. A lower dip location indicates a stronger AC effect. The geometry types corresponding to each of the symbols used (circle, square, diamond, star, triangle) are shown in the SI, along with an equivalent plot to this one, with a one-to-one color map of each geometry and its corresponding HCACF integral. The bottom right inset corresponds to a zoom-in on the region of the left corner of the plot. The top-left inset compares the actual (fractional) thermal conductivity to the thermal conductivity predicted for $\mathrm{C = (2-\alpha)/(2\alpha)}$, i.e., under the assumption of standard scattering conditions.}\label{fig:fig3}
\end{SCfigure}

We have fitted $\mathrm{C}$ in Eq.~\eqref{eqn:model} to nearly 40 distinct porous geometries with a uniform $d$ spacing. Note that the total thermal conductivity is obtained by integrating $\Lambda_\mathrm{eff}$ over all $\lambda$. $\mathrm{C}^{-1}$ is plotted against $2\alpha/\left(2-\alpha\right)$ in Fig.~\ref{fig:fig3}, with the data points colored by the magnitude of the HCACF dip if one exists. At low $\alpha$ the plots with large HCACF dip have a $\mathrm{C}^{-1}$ that is suppressed significantly below that which is predicted by the pore geometry ($\mathrm{C}^{-1} < 1$) indicating phonon MFP super-suppression. Super-suppression of the phonon MFP is also observed in the ray tracing simulations for all four scattering models, as is plotted in the SI, in Fig.~\ref{fig:SI-super-suppression-MC}, and discussed in subsequent sections. In all cases, the larger supper-suppression occurs in conjunction with a large HCACF dip.

\subsection{Ray Tracing Model of the  HCACF}\label{sec:theory-bulk-conductivity}

It is difficult to infer the phonon scattering behavior that gives rise to the anticorrelation in the HCACF directly from the shape of the HCACF curve. To circumvent this difficulty, we have developed {an equilibrium} Monte Carlo (MC) phonon ray tracing model which allows us to test how different scattering behaviors that might occur at pores would be manifest in the HCACF, allowing us to seek signatures of different scattering phenomena in the HCACF obtained from the MD simulations.

Phonon ray tracing models have been used to great effect to model thermal transport~\cite{mazumder2001monte, chakraborty2018monte, chakraborty2019thermal, wolf2014thermal, jean2014monte, wang2011thermal}, and in many cases to also identify signatures of ballistic transport in a variety of nanoscale geometries~\cite{lee2017investigation}. 
{The MC ray tracing approach that we use here, however, is significantly different from the typical approach used in these works. Rather than simulating a system under an imposed temperature gradient, as is typically the case (as for example in Refs.\cite{yang2017thermal, PhysRevB.104.195303}), we simulate a system \emph{at equilibrium}, and use the Green Kubo method to compute the thermal conductivity indirectly from the HCACF, exactly the approach used in equilibrium MD simulations. To the best of our knowledge, calculation of the HCACF from a ray tracing model is completely novel. Non-equilibrium MC simulations are by necessity quite computationally complex as in order to satisfy detailed balance they require one to keep track of the local temperature everywhere in the spatial domain being modeled. Our simulations are much simpler. We simulate the birth, flight and annihilation of phonons in a periodic real space domain, that being at equilibrium, has a uniform temperature.} 

{In the next section, we first lay out the theoretical relationship between the total HCACF and the HCACF from individual phonon rays in a pristine bulk crystal. This provides the theoretical foundation for Monte Carlo simulations in the sections that follow in which we compute the total HCACF in nanoporous systems from the  HCACF of individual phonons undergoing scattering.}

\subsubsection{Stochastic Model of Heat Current Fluctuations in Bulk Crystal}\label{sec:Stochastic-Model-of-Heat-Current}

In the MD simulations the thermal conductivity is computed from the natural fluctuation in heat current for a system at equilibrium using the Green-Kubo (GK) formalism and the expression
\begin{equation}
     \mathbf{\kappa} = \frac{V}{k_BT^2}\int_{0}^{\infty } d\tau \left\langle \mathbf{J}(t) \otimes  \mathbf{J}(t+\tau)\right\rangle,
\label{eq:kappa_GK}
\end{equation}
where $\mathbf{J}(t)$ is the total instantaneous heat flux at time $t$, in the volume of the material, $V$, at temperature $T$. The heat current autocorrelation function (HCACF), $\left\langle \mathbf{J}(t) \otimes  \mathbf{J}(t+\tau)\right \rangle$, is the time-averaged dyadic of the heat fluxes an interval $\tau$ apart -- it  measures the size and longevity of thermal fluctuations in the heat flux of a system in equilibrium. 

The GK approach and the evaluation of the HCACF with MD is described in detail in SI Sec.~\ref{sec:computational-methods}, but here we derive an analytic expression for the HCACF in a bulk crystal. The total instantaneous heat flux that appears in Eq.~\eqref{eq:kappa_GK} is the sum of the instantaneous heat flux fluctuations from all phonon modes as: 
\begin{equation}
    \mathbf{J}(t) = \sum_{\mathbf{k}p} \mathbf{J}_{\mathbf{k}p}(t) 
    = \sum_{\mathbf{k}p} \left( n_{\mathbf{k}p}(t) -   \left\langle n_{\mathbf{k}p} \right\rangle  \right) \frac{\hbar \omega_{\mathbf{k}p} \mathbf{v}_{\mathbf{k}p}}{V}.
\end{equation}
Here $n_{\mathbf{k}p}(t)$ and $\left\langle n_{\mathbf{k}p} \right\rangle$ are the instantaneous and average occupancy of the phonon mode with wave vector $\mathbf{k}$ and polarization $p$. The terms $\omega_{\mathbf{k}p}$ and $\mathbf{v}_{\mathbf{k}p}$ are the mode's angular frequency and group velocity. If the occupancy fluctuations in one mode are uncorrelated with the fluctuations in the other modes, it is shown in Appx.~\ref{sec:apndx-theory-bulk-conductivity} that the total HCACF can be written as the sum of the occupancy auto-correlation function for each mode individually as:
\begin{equation}
    \left\langle \mathbf{J}(t) \otimes  \mathbf{J}(t+\tau)\right\rangle = 
    \sum_{\mathbf{k}p} \left ( \frac{\hbar \omega_{\mathbf{k}p}}{V} \right )^2 \mathbf{v}^2_{\mathbf{k}p} \left\langle  \left( n_{\mathbf{k}p}(t) -   \left\langle n_{\mathbf{k}p} \right\rangle  \right) \left( n_{\mathbf{k}p}(t+\tau) -   \left\langle n_{\mathbf{k}p} \right\rangle  \right)    \right\rangle,
\label{eq:J(t)-stochastic}
\end{equation}
where the shorthand notation $\mathbf{v}^2_{\mathbf{k}p} = \left(\mathbf{v}_{\mathbf{k}p} \otimes  \mathbf{v}_{\mathbf{k}p} \right)$ has been used for the tensor product of the group velocity.

\begin{figure*}[t]
    \centering
    \subfloat{\includegraphics[width=1.0\textwidth]{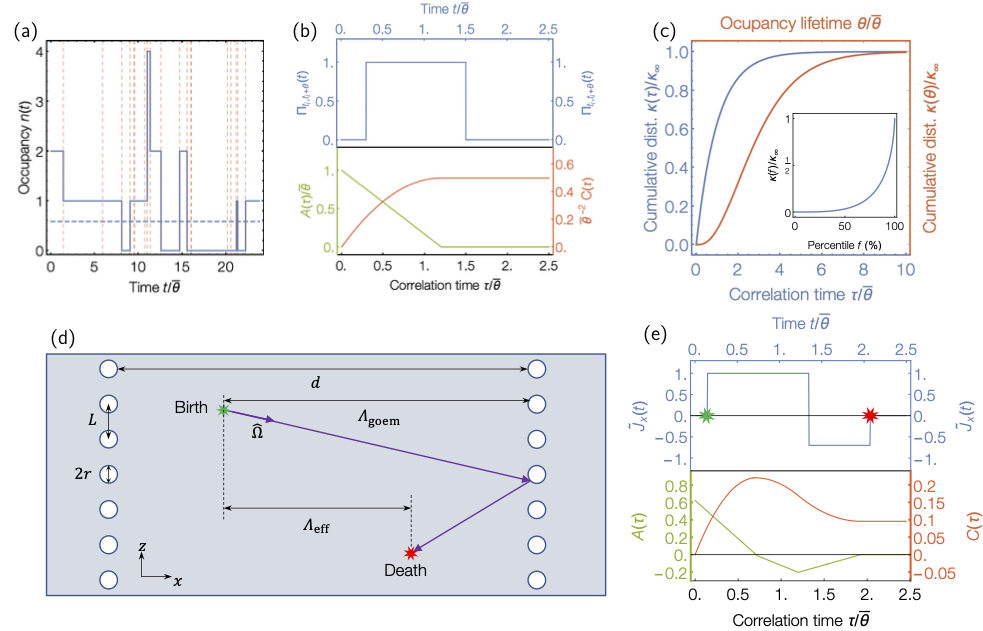}}
    \caption{(a) Example of the random occupancy fluctuations in a phonon mode with dimensionless frequency $\widetilde{\omega}=1.0$ and mean scattering lifetime $\overline{\theta}$. (b) Example of a single-occupancy fluctuation in time (top pane) and its autocorrelation (green) and integral (red) in the bottom pane. (c) The plot of the cumulative thermal conductivity distributions for bulk materials with a gray phonon population. The blue plot is the distribution over correlation time, $\tau$ (the integral of the HCACF) and the red line is cumulative conductivity over the distribution of random phonon lifetime, $\theta$. The inset plot shows the distribution over occupancy events showing that most of the heat is carried by only a small number of, particularly long-lived phonons. (d) Schematic of a phonon wavepacket’s trajectory in nanoporous Si. In this example the wavepacket is spawned as the result of anharmonic phonon-phonon interactions at the green starburst, traveling initially along direction $\hat{\Omega}$, and is scattered elastically from pores before its annihilation through another phonon-phonon interaction at the red starburst. Because of this backscattering, the phonon only carries heat forward by a net distance $\Lambda_{\mathrm{eff}}$ which is considerably shorter than the geometric free-path, $\Lambda_{\mathrm{geom}}$, that the phonon travels before colliding with the pores. The heat flux and HCACF contribution from this ray are shown in panel (e). The top pane shows the contribution to heat flux along the $x$ direction (blue). The bottom pane shows the corresponding HCACF and its integral plotted in green and red, respectively.
    }   
    \label{fig:stochastic-model-bulk}
\end{figure*}

The occupancy $n_{\mathbf{k}p}(t)$ of a phonon mode will be a random stepped function in time as shown in Fig.~\ref{fig:stochastic-model-bulk}(a). The phonon mode holds a constant excitation for a random interval $\theta$ between anharmonic interactions with other phonon modes, which lead to scattering and a reset of the mode’s excitation. It is worth clarifying here that the nomenclature used for the three different quantities have dimensions of time. The letter $t$ is used to denote a point in time (clock time), whereas the correlation time, which is the interval or delay between two points in time is denoted with $\tau$. The lifetime of an individual phonon is denoted by the symbol $\theta$, and the average lifetime of all phonons by $\overline{\theta}$.

At any time $t$ the probability of finding the mode in the $\mathrm{n^{th}}$ state of occupancy where $P_n$ is the probability distribution for the canonical ensemble, is:
\begin{equation}
    P_n = e^{-n\widetilde{\omega}} \left(1 - e^{-\widetilde{\omega}} \right),
    \label{eq:Pn_txt}
\end{equation}
with $\widetilde{\omega}=\frac{\hbar\omega}{k_BT}$ being the mode's dimensionless frequency for the given temperature. The modal heat current $\mathbf{J}_{\mathbf{k}p}(t)$ can be expressed as a sequence of boxcar functions. If we assume that the occupancy value of a mode excitation before and after a scattering event are also uncorrelated, then the mode's HCACF is simply the average of the autocorrelation function of individual boxcar functions weighted by the probability of their amplitude and duration. Figure~\ref{fig:stochastic-model-bulk}(b) shows a single boxcar function, $\Pi_{t_0,t_0+\theta}\left(\theta\right)$ for a fluctuation that starts at time $t_0$ and lives for lifetime $\theta$. Shown in the lower panel is its non-normalized ACF, $A(\tau,\theta)$, which is just a ramp function, and the integral of the ACF, $\mathrm{C}\left(\tau,\theta\right)$. Analytic expressions for $A$ and $\mathrm{C}$ are given in Appx.~\ref{sec:apndx-theory-bulk-conductivity}. If we further assume that scattering processes are random so that the phonon lifetimes are drawn from a Poisson distribution of waiting times with an average wait time $\bar{\theta}$ that is independent of occupancy, then the averaging of the occupancy ACF in Eq.~\eqref{eq:J(t)-stochastic} can be performed analytically. After substitution into Eq.~\eqref{eq:kappa_GK}, and some  manipulation (see Appx.~\ref{sec:apndx-theory-bulk-conductivity}), the expression for the cumulative thermal conductivity can be derived
\begin{equation}
\mathbf{\kappa}(\tau)= \sum_{\mathbf{k}p} \left( \frac{\hbar \omega_{\mathbf{k}p}}{V} \frac{d \langle n_{\mathbf{k}p}\rangle}{dT} \right)  v^2_{\mathbf{k}p} \overline{\theta}_{\mathbf{k}p} \left(1 - e^{-\tau/\overline{\theta}_{\mathbf{k}p}} \right).
\label{eq:kappa-bulk-stochastic}
\end{equation}
The first term in parenthesis is the mode’s volumetric specific heat, meaning that if the phonon modes all have the same scattering rate and group velocity, assuming cubic symmetry, the first element of the thermal conductivity tensor reduces to the well-known result from kinetic theory:
\begin{equation}
    \kappa_{xx}(\infty) = \frac{1}{3} C_v v_g \lambda,
\end{equation}
where $C_v$, $v_g$, and $\lambda$ are the systems' volumetric specific heat, average group velocity, and mean free path. This demonstrates that, if there are no hydrodynamic effects in the phonon scattering, we can relate the macroscopic HCACF of a system to the autocorrelation function of individual excitations of individual phonon modes. 

In MD simulations, the HCACF is computed from the collective excitation of all phonon modes. The result above shows that we can interpret the macroscopic HCACF by examining the contribution from individual phonon modes and scattering processes alone, without the need to consider a collective excitation. This provides the formal theoretical footing for the ray tracing model that follows, but before moving on to the ray tracing model, we make note of several insights that can be obtained from the analysis above. 
 
The HCACF obtained from the MD simulations includes contributions from acoustic and optical phonons with a range of different frequencies, MFPs, group velocities, and directions of travel. It can be tempting to interpret the integrated HCACF as a cumulative distribution function of phonon contributions to thermal conductivity, but this is not the case and masks phenomena that arise from the stochastic nature of phonon scattering. This can be understood by considering the stochastic model for thermal conductivity of a bulk grey medium, where all phonons have the same \emph{average} lifetime $\overline{\theta}$, but the lifetime $\theta$ of every individual phonon is random. The HCACF in Eq.~\eqref{eq:ave} decays exponentially over a correlation time equal to $\overline{\theta}$, so the cumulative thermal conductivity distribution over \emph{correlation} time $\tau$ is
\begin{equation}\label{eq:kappa_ratio_1}
    \frac{\kappa(\tau)}{\kappa_{\infty}} = 1 - e^{-\tau/\overline{\theta}},
\end{equation}
which is plotted in blue in Fig.~\ref{fig:stochastic-model-bulk}(c). However, not all phonons contribute equally to this. As can be seen from Eq.~\eqref{eq:C}, the contribution that an occupancy fluctuation makes to the thermal conductivity is proportional to its lifetime squared, meaning that the rare long-lived occupancy fluctuations in the tail of the Poisson distribution contribute considerably more to heat transport than the more numerous short-lived fluctuations. Using this we can compute the cumulative thermal conductivity distribution over phonon \emph{lifetimes}, $\theta$, as
\begin{equation}\label{eq:kappa_ratio_2_txt}
    \frac{\kappa(\theta)}{\kappa_{\infty}} = \frac{\int_{0}^{\theta}d\theta' \theta'^2 P_{\overline{\theta}}(\theta')}{\int_{0}^{\infty}d\theta' \theta'^2 P_{\overline{\theta}}(\theta')} = 1- \frac{1}{2} \left( 2 + \frac{\theta}{~\overline{\theta}~} \left( 2 + \frac{\theta}{~\overline{\theta}~}\right) \right)e^{-\theta/\overline{\theta}}.
\end{equation}
This cumulative conductivity with increasing phonon lifetime is plotted in red in Fig.~\ref{fig:stochastic-model-bulk}(c). It can be seen in this plot that short-lived fluctuations (small $\theta$) contribute very little to the thermal conductivity while the cumulative conductivity distribution in $\tau$ rises rapidly at small correlation times. This illustrates the conceptual difference between fluctuation duration $\theta$ and correlation time $\tau$; the change in $\mathbf{\kappa}(\tau)$ at correlation time $\tau$ includes contributions from all fluctuations with $\theta >\tau$.  Note, that it is common in heat transport texts to see plots of the cumulative contribution to thermal conductivity over the distribution of MFPs from phonons with a spectrum of frequencies  $\mathbf{\kappa}(\lambda)$, and so we take pains here to point out that this is different from the result in Eq.~\eqref{eq:kappa_ratio_2_txt}, which comes from the shot noise in a population of phonons with the \emph{same} frequency.  However, the effect of this noise is significant, with a large contribution to the total conductivity coming from a small number of phonons that travel ballistically over distances many times longer than the \emph{mean free path} before scattering. This point is further emphasized by computing the cumulative conductivity distribution over the fraction $f$ of occupancy fluctuations ranked in ascending order of their longevity,
\begin{equation}
    \frac{\kappa(f)}{\kappa_{\infty}} = f + (1-f) \ln{(1-f)}\left( 1 - \frac{1}{2} \ln{(1-f)}\right),
\end{equation}
which is plotted inset in Fig.~\ref{fig:stochastic-model-bulk}(c). More than 50\% of the heat transport is carried by just 7\% of phonons that survive for more than 2.7 times the average phonon lifetime $\overline{\theta}$. It can also be seen that 80\% of the heat current comes from fewer than 20\% of the phonon mode occupancy fluctuations---an instance of Pareto’s 80-20 rule, but one that does not arise from a power-law distribution of flight distances as is the case for a L\'evy flight. This observation has important ramifications for deterministic simulations of the Boltzmann transport equation (BTE) for phonons, implying that to correctly predict the heat conduction due to phonon transport in the nanostructured material, one must smear the intrinsic distribution of average phonon lifetimes by the Poisson distribution—a practice that is often overlooked in frequency-dependent and multi-grey BTE simulations~\cite{harter2019prediction, romano2021openbte}.

\subsubsection{Monte Carlo Ray Tracing Model for Correlated Scattering}\label{sec:Monte-Carlo}

To isolate the effects that arise from back-scattering, we study a simpler model system consisting of a grey population of phonons in which we assume that all phonon modes have the same frequency $\omega$, group velocity $v_g$, and \emph{mean} intrinsic scattering lifetime $\overline{\theta}$. Similar to the geometries of interest, the scattering centers are laid out in rows as shown in Fig.~\ref{fig:stochastic-model-bulk}(d). The rows have a spacing $d$, and within each row, the pores have radius $r$ and spacing $L$. The derivation for bulk thermal conductivity above shows that the HCACF can be constructed by considering each correlated heat carrier event in isolation and then averaging their contribution to the total HCACF. For a wavepacket scattered elastically from one mode (with wave vector $\mathbf{k}$ and polarization $p$) into another mode ($\mathbf{k'}p'$) by an interface, one must consider the flight of both the incident wavepacket and the scattered wavepacket together as the occupancy fluctuations in $\mathbf{k}p$ and $\mathbf{k'}p'$ modes are now (anti)correlated. This is true even in the case of diffuse scattering, where the choice of the scattered $\mathbf{k'}$ mode is independent of the incident mode, as after scattering the \emph{sign} of the velocity component perpendicular to the interface is reversed, making the heat fluxes before and after scattering anticorrelated. As we now must consider the sequential occupancy of two or more modes, rather than compute the occupancy auto- and cross-correlations of the modes, we instead consider the heat current from individual packets of lattice vibration, starting from their birth, and following them as they are scattered elastically through a series of different phonon modes, up until the uncorrelated phonon-phonon scattering event that causes their annihilation. An example of such a trajectory is shown in Fig.~\ref{fig:stochastic-model-bulk}(d) with the resulting heat flux along $x$ and its autocorrelation function shown in Fig.~\ref{fig:stochastic-model-bulk}(e). The ray diagram in Fig.~\ref{fig:stochastic-model-bulk}(d) also illustrates the concept of mean free path super-suppression: Elastic backscattering of phonons before they thermalize with the phonon bath means that the effective distance over which heat is carried, $\Lambda_\mathrm{eff}$, is considerably shorter than the line-of-sight distance imposed by the geometry, $\Lambda_\mathrm{geom}$. The result is that the total mean free path is suppressed by more than Matthiessen's rule would predict from the geometric value $\Lambda_\mathrm{geom}$, and so we say that the mean free path is \emph{super-suppressed}.

In using this conceptually subtly different approach of following energy packets (phonons) rather than occupancy fluctuations, we can apply some of the insights from the derivation above to write the autocorrelation function of the heat current fluctuations along the $x$ direction as the average of the heat flux autocorrelation functions of wavepackets with the unit magnitude as
\begin{equation}
    \langle J_x(t)J_x(t+\tau) \rangle = \frac{D}{V \overline{\theta}}\frac{e^{\widetilde{\omega}}}{(e^{\widetilde{\omega}}-1)^2}(\hbar \omega v_g)^2 \langle A_{xx}(\tau, \theta, \mathbf{r}, \widehat{\Omega}) \rangle,
\end{equation}
where $D$ is the density of states (the number of phonon modes per unit volume), and $A_{xx}(\tau, \theta, \mathbf{r}, \widehat{\Omega})$ is the autocorrelation function of the heat flux along $x$ created by a unit wavepacket that was born at location $\mathbf{r}$, traveled initially along direction $\hat{\Omega}$, and lived for duration $\theta$, before annihilation into the phonon bath. The total thermal conductivity reduction can thus be computed as:
\begin{equation}\label{eq:k_kbulk}
    \frac{\kappa(\tau)}{\kappa_{\mathrm{bulk}}} = \frac{3}{~\overline{\theta}^2~} \int_0^{\tau}d\tau' \langle A_{xx}(\tau', \theta, \mathbf{r},\widehat{\Omega}) \rangle = \frac{3}{~\overline{\theta}^2~} \langle C_{xx}(\tau,\theta, \mathbf{r},\widehat{\Omega}) \rangle.
\end{equation}
As this population of phonons is in a volume that includes physical scattering centers such as the pores, in addition to averaging over the phonon lifetimes $\theta$, the average $\langle A_{xx}(\tau', \theta, \mathbf{r},\widehat{\Omega}) \rangle$ is also taken over the spatial domain $\mathbf{r}$, the phonon modes $\widehat{\Omega}$, and the various possibilities for the reflected wavepacket at each correlated scattering event. Rather than performing the average in Eq.~\eqref{eq:k_kbulk} analytically, we average using Monte Carlo sampling---tracing the trajectory of wavepackets as they collide with pores and are scattered off into new directions. In this scheme we took the average HCACF from 20,000 randomly sampled wavepacket trajectories; each beginning from a randomly selected starting point, initial direction, and lifetime, the latter drawn from the Poisson distribution. The single wavepacket HCACFs in the $x$, $y$, and $z$ directions were computed numerically out to a correlation time of 20 times the mean scattering time $\overline{\theta}$ to prevent truncation of contributions from long-lived phonons in the tails of the Poisson distribution. This averaging was sufficient to reduce the uncertainty in the computed values of $\kappa_{\mathrm{bulk}}$ to $< 3\%$.

To elucidate how the different features in the HCACF are related to the details of how phonons behave when they encounter a pore, we compare four different phenomenological models for scattering in the ray tracing simulations. These were derived from two different scattering center geometries: an array of permeable planer \emph{walls}, and a series of \emph{palisades} of cylindrical pores. For each geometry we have examined the effect of both specular and diffuse phonon scattering by the obstacles, giving us four scattering models in total. Both the wall and palisade models have a single geometric degree of freedom, $\alpha$ that is related to the probability that phonons are not scattered when they encounter the wall/palisade, and the details of the backscattering behavior in each model are described in detail below. 

\paragraph{Diffuse and specular wall scattering models} 
The "wall" scattering models make drastic simplification of the pores in the MD simulation by replacing them with a series of parallel planar interfaces perpendicular to the $x$-axis. The spacing, $d$, between planes is described in terms of the Knudsen number $K_n = \frac{v_g \overline{\theta}}{d}$. When a phonon encounters an interface perpendicular to its transport direction, there is a finite probability $1-\alpha$ that it will pass through with its trajectory unaltered, otherwise, the phonon is reflected. For the specular model, reflection involves only flipping the direction of the $x$-component of the velocity. In the diffuse model, a new random directory is chosen in the half-space perpendicular to the reflection plane, so that the $x$-component of the velocity of the scattered phonon has the opposite sign from the incident phonon. Simulations were performed sweeping $\alpha$ from 0 to 1 and $\log \left[K_n\right]$ from -1 to 1. Although we refer to this model as the interface model, it is a reasonable representation of scattering from rectangular pores.

\paragraph{Diffuse and specular palisade scattering models} 
With the palisade model, we aim to more closely mimic the pore geometry simulated in the MD simulations. We assume that the material contains rows of cylindrical pores, each with radius $r$ and aligned with their axis parallel to the $y$-direction. The pores are arrayed in the $z$-direction with spacing $L$ to form a palisade fence. The geometry of the palisade is parametrized with $\alpha = \frac{2r}{L}$, which is the scattering probability for phonons that meet the palisade head-on. The probability that a phonon passes between the cylinders in the palisade depends on the gap between cylinders that is visible to the phonon as it travels towards the palisade. This means that the transmission probability depends on $\alpha$ and the angle of incidence of the phonons, and there will be a range of grazing angles for which the gaps are shadowed and so the transmission probability is zero. The full description of how the angular-dependent transmission probability is computed is provided in the SI. When a phonon’s $x$ position lies on a palisade a pseudorandom number generator is used to decide if the phonon passes through the barrier unscattered or if it strikes on one of the pillars.

If the phonon is scattered, the determination of the outgoing scattered ray direction is similar to the procedures used in the wall models. For the specular palisade model, the normal angle for the reflecting surface is selected from the distribution function described in the SI, and the phonon is set on a new trajectory based on specular reflection from this surface. Note that this mean-field model of scattering does not resolve the physical size of the pores and the possibility of multiple scattering between neighboring cylinders within a palisade is not considered. The model for diffuse scattering from cylindrical pores uses the same procedure to stochastically select the normal at the point on the surface of the cylinder that the phonon strikes. A new random direction is then selected for the phonon in the half-space defined by this normal vector.

\begin{figure*}[t]
    \centering
    \includegraphics[width=1\textwidth]{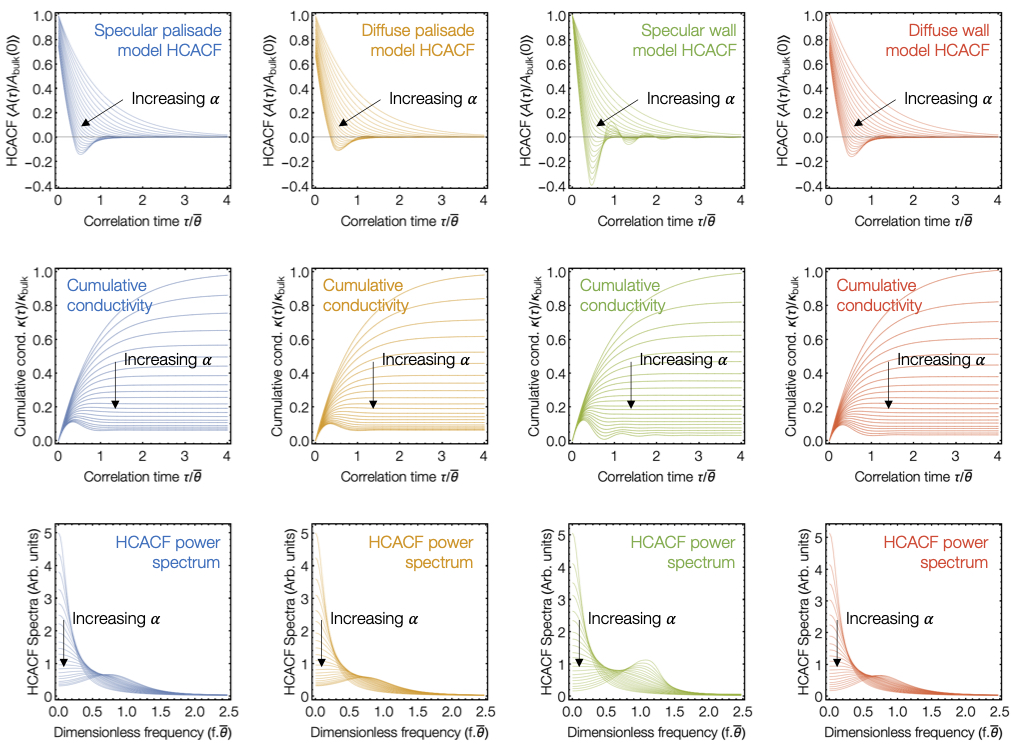}
    \caption{Plots of the HCACF (top row), and corresponding cumulative thermal conductivity (bottom row), for the four scattering models. The specular and diffuse palisade models are plots in blue (a, e \& i) and gold (b, f \& j), respectively, while specular and diffuse wall models are plotted in green (c, g \& k) and red (d, h \& l), respectively. All plots are for simulations with $K_n=2.5$, and the geometric scattering probability $\alpha$ is swept from 0 to 1. In the top row, the HCACF is plotted normalized by $A_{\mathrm{bulk}}(0)=1/3 (v_g \tau_o )^2$, the initial value of the HCACF in the bulk crystal. The middle row shows the cumulative thermal conductivity normalized by the thermal conductivity of the bulk crystal and the bottom row shows the power spectrum of the normalized HCACF.}
    \label{fig:fig7}
\end{figure*}

The HCACF and corresponding cumulative conductivity obtained with the four different scattering models is shown in Fig.~\ref{fig:fig7} for a wide range of the geometric parameter $\alpha$, which is a surrogate for the pore diameter to spacing ratio $2r/L$. A complementary set of plots showing the change in HCACF with increasing $K_n$ is included in the SI (Fig.~\ref{fig:figA2}). It can be seen from these that all four scattering models can generate a strongly anticorrelated HCACF when $\alpha$ approaches 1 and the Knudsen number $K_n$ becomes large -- with all models achieving a magnitude of AC and thermal conductivity reduction comparable with that seen in the MD simulations. In these plots, the trends in the HCACF for the two palisade scattering models look very similar, but there are marked differences between the palisades scattering models and the behavior observed for the specular and diffuse walls. 

The most striking difference between models in Fig.~\ref{fig:fig7} is the number and position of the HCACF minima.  When the Knudsen number is much larger than 1, most of the phonons will experience multiple reflections back and forth between the walls/palisades. It can be seen in Fig.~\ref{fig:fig7}(c) that under these conditions specular permeable walls produce a ringing anticorrelation with a sequence of dips (intervals of anticorrelation) in the HCACF. However, under the same multiple scattering conditions, the diffuse wall model and the two palisade models only produce a single dip in the HCACF. This latter behavior implies that although phonon wavepackets are scattered multiple times they lose the memory of their direction of travel (momentum) after the second scattering event---even in the case of specular scattering from cylindrical reflectors. In the MD simulations, only one dip in the HCACF is ever observed, suggesting no correlation in the momentum of the two scattering events.

\begin{figure*}[ht]
    \centering
    \includegraphics[width=1.0\textwidth]{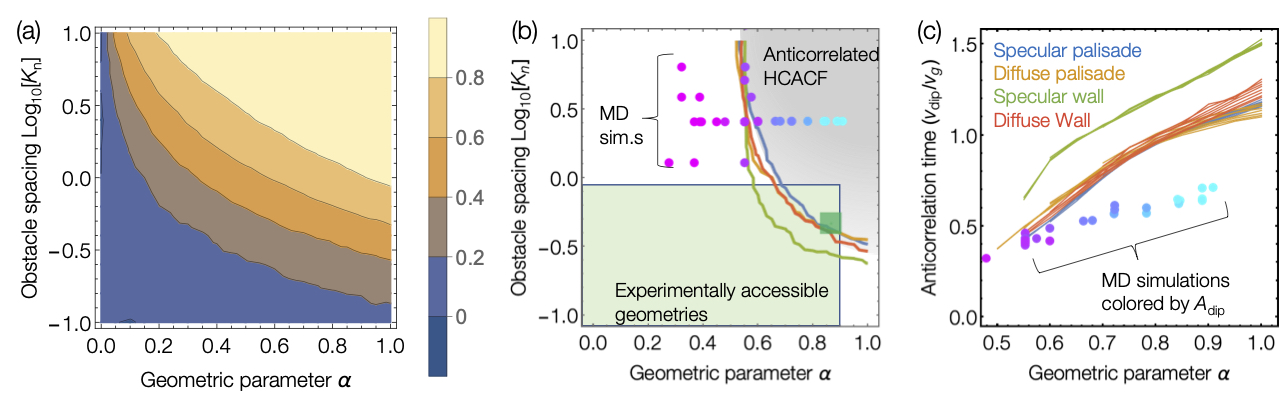}
    \caption{Plot (a) shows a contour map of the thermal conductivity reduction, $1-\kappa/\kappa_\mathrm{bulk}$, predicted by the ray tracing model for the diffuse palisade model as a function of the $\alpha$ and $K_n$. Plot (b) show the phase map for the conditions of $\alpha$ and $K_n$ under which a detectable anticorrelation is observed in the HCACF for the four scattering models. The grey shaded region of the HCACF is anticorrelated, and the blue, gold, green and red lines are the boundaries for the onset of observable anticorrelation for the specular and diffuse palisade models and the specular and diffuse wall models, respectively.  Interestingly, the boundaries for the onset of AC do not follow the contours of constant thermal conductivity shown in plot (a). The region shaded in green in (b) represents geometries that are experimentally accessible using fabrication methods described in Ref.~\cite{tang2010holey}, with the green square representing pores with 65 nm radius arranged in a rectangular grid with long and short spacings of 400 and 150 nm showing that anticorrelation regime should be experimentally accessible. The circular points show the geometries of the molecular dynamics simulations in this study, color-coded by the depth of their ACF dip using the same color scheme as in Fig.~\ref{fig:fig3}. The onset of the AC in MD models lines up with the prediction from the ray tracing model.  The boundaries for the diffuse wall and the two palisades’ models are within the noise of one another, but the specular wall is distinctly different. Plot (c) shows the reciprocal of time to the maximum anticorrelation, $\tau_\mathrm{dip}^{-1}$, expressed in terms of the effective velocity $v_\mathrm{dip}=d \tau_\mathrm{dip}^{-1}$, normalized by the group velocity, for all the simulations performed. The solid lines are for the four ray tracing models and use the same color coding as in (b). Each line corresponds to a sweep of $\alpha$ for different $K_n$. It can be seen that $v_\mathrm{dip}$ is largely independent of the obstacle spacing $d$ for all models, and again the specular wall model is markedly different from the other three models. In all cases $v_\mathrm{dip}$ increases with increasing scattering probability, $\alpha$. The circular markers show the results from the MD simulations performed in this study color-coded according to the strength of the anticorrelation with the same color scheme as in (b). The ray tracing model overestimates $v_\mathrm{dip}$ (underestimates $\tau_\mathrm{dip}$) and $\tau_\mathrm{dip}$ from MD simulations is much less sensitive to $\alpha$.}
    \label{fig:fig8}
\end{figure*}

A second significant difference between the scattering models is that the different modes of scattering considered give rise to a different dependence of the HCACF on the geometric arrangement of scattering centers as parameterized with $\alpha$ and $K_n$. Fig.~\ref{fig:fig8}(a) shows a contour map of the thermal conductivity reduction, $1-\kappa/\kappa_\mathrm{bulk}$, as a function of the system geometry described by $\alpha$ and $K_n$ for the diffuse palisade model, and is representative of the same plot in the other three models. Fig.~\ref{fig:fig8}(b) maps for the same range of $\alpha$ and $K_n$ the conditions under which backscattering produces a discernible anticorrelation (dip) in the HCACF (the grey shaded region). The four colored lines mark the precise boundary for the onset anticorrelated behavior of four scattering models using the same color coding as in figure~\ref{fig:fig7}. The boundaries for the two palisade models and the diffuse wall are very similar (within the numerical noise in the calculation); however, specular wall scattering can produce an anticorrelated HCACF with much larger obstacle spacing (lower $K_n$) than is required for the other scattering modes. Interestingly, none of the boundaries for the onset of anticorrelated behavior in Fig.~\ref{fig:fig8}(b) follow the thermal conductivity contours in Fig.~\ref{fig:fig8}(a). Also shown in Fig.~\ref{fig:fig8}(b) are the geometries of the systems simulated with MD in this work. These have been color-coded by the magnitude of their HCACF dip, and the onset of anticorrelated behavior in the MD simulations matches that for all four ray tracing models. The pale green region in Fig.~\ref{fig:fig8}(b) indicates the domain of geometries that are accessible experimentally using fabrication and patterning techniques such as those described in Ref.~\cite{tang2010holey}. This domain overlaps with the domain where anticorrelated behavior is expected, and thus we predict that it is possible to engineer nanostructures that exhibit strongly anticorrelated heat transport. For example, the darker green square in this overlapping region corresponds to an array of 130 nm diameter pores arranged in a rectangular grid with long and short spacings of 400, and 150 nm.

One difference between the MD and MC models is the interval to the dip in the HCACF, $\tau_\mathrm{dip}$ and its sensitivity to the pore spacing ratio $\alpha$. Fig.~\ref{fig:fig8}(c) plots the inverse time interval to the dip, $\tau_{\mathrm{dip}}^{-1}$, expressed as an effective velocity $v_\mathrm{dip}=d/\tau_{\mathrm{dip}}$ to normalize for different separations between the ranks of walls/palisades. This is plotted for a wide range of geometries varying both $\alpha$ and $K_n$. It can be seen from this plot that the HCACF dip in MD simulations occurs at longer correlation times than in the ray tracing simulations. This is likely because the MD simulations contain the full phonon spectrum and so the HCACF included contributions from diffusive and ballistic phonons. From the MD simulations in our previously reported work we found $v_\mathrm{dip}/v_g\sim0.6$~\cite{PhysRevB.102.205405}. The sets of MC simulations with a grey phonon population show that there is no fundamental reason why $v_\mathrm{dip}$ is constant and with the wider range of MD simulated geometries in this work, we see that it is not.

\subsubsection{Signatures of scattering behavior in the HCACF}

\begin{figure*}[h!]
    \centering
    \includegraphics[width=1\textwidth]{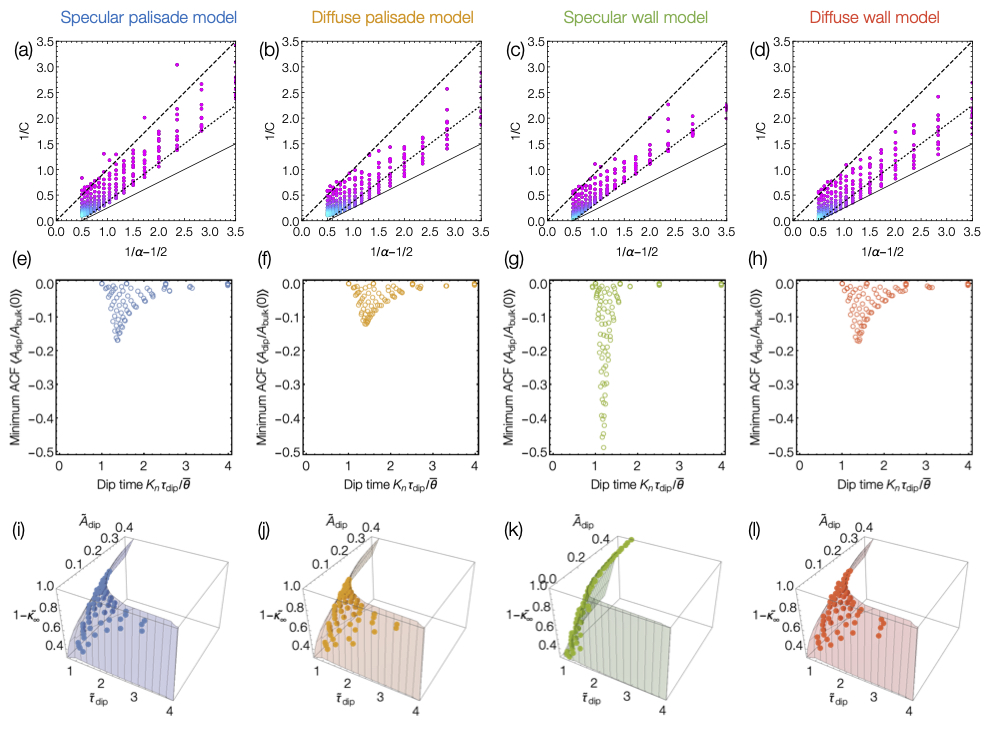}
    \caption{The plots in the top row (a--d) show the scattering strength constant, $\mathrm{C}$, obtained from the ray tracing simulations with specular (a) and diffuse (b) palisade models of scattering and with the specular (c) and diffuse (d) wall scattering models. In each the markers are colored according to the depth of the HCACF dip using the same color scheme as Fig.~\ref{fig:fig3}. On each the dashed line indicates the $\nicefrac{1}{\alpha}-\nicefrac{1}{2}$ line, the dotted line is for $\nicefrac{1}{2}\left(\nicefrac{1}{\alpha}-1\right)$ and the solid line for $\nicefrac{3}{4}\left(\nicefrac{1}{\alpha}-1\right)$. The plots in the middle row show the location (time and depth) of the minimum of the HCACF for all simulations with detectable anticorrelation. The anticorrelation time $\tau_{dip}$ is scaled by the Knudsen number and normalized by the mean phonon lifetime. The specular and diffuse palisade and specular and diffuse wall scattering models are plotted in blue, gold, green, and red respectively (e--h). The plots in the bottom row show the time and depth of the HCACF minimum, along with the corresponding thermal conductivity reduction $1-\tilde{\kappa}_\infty$ with the same color coding as in the top row. For each scattering model, all the points lie on a single manifold meaning that the location and depth of the anticorrelation are sufficient to predict the final thermal conductivity. The translucent surface in plots (i--l) shows the fit through these points as described in the main text. The manifold is different for each scattering model, providing a means of distinguishing the scattering behavior present in the MD simulations.}
    \label{fig:fig9}
\end{figure*}

To examine how different modes of backscattering suppress the effective phonon mean free path, in Fig.~\ref{fig:fig9}(a--d) we plot for the four different sets of ray tracing simulations the scattering strength constant, $\mathrm{C}$, \emph{vs.} the dependence on $\alpha$ predicted by Matthiessen's rule, $1/\alpha-1/2$. These follow the format of Fig.~\ref{fig:fig3}, but instead of also drawing the line $1/\alpha-1$ as a guide to the the eye we have plotted guiding lines for $\frac{3}{4}\left(1/\alpha-1\right)$  and $\frac{1}{2}\left(1/\alpha/-1\right)$. When the HCACF is strongly anticorrelated the scattering strength $\mathrm{C}$ is considerably larger than that predicted by Matthiessen's rule -- the effective phonon mean free path is \emph{super-suppressed} below the mean line-of-sight distance imposed by the geometry. Interestingly we also see that the limiting behavior with strong backscattering is different in the models with diffuse scattering than those with specular scattering and that, surprisingly, diffuse scattering yields a larger super-suppression of MFP than specular scattering.

A final and unexpected finding from this work is that there exists a hidden relationship between the position and depth of the HCACF, which together define the ultimate thermal conductivity reduction. As can be seen in Fig.~\ref{fig:fig9}(e--h), when looking at many simulation results plotted together, there appears to be qualitative differences in the position and depth of the HCACF anticorrelation dip for the four scattering models. To examine this more quantitatively we characterize each HCACF trace by three dimensionless descriptors: the anticorrelation time scaled by the Knudsen number $\tilde{\tau}_\mathrm{dip} = K_n\tau_\mathrm{dip}/\overline{\theta}$; the depth of the dip $\tilde{A}_\mathrm{dip} = A\left(\tau_\mathrm{dip}\right)/A_\mathrm{bulk}(0)$ normalized by the $\tau=0$ value of the HCACF for the bulk medium; and the normalized thermal conductivity reduction $1-\tilde{\kappa}_\infty$ where $\tilde{\kappa}_\infty=\kappa(\infty)/\kappa_\mathrm{bulk}$. Figure~\ref{fig:fig9}(e--h) shows scatter plots of $\tilde{\tau}_\mathrm{dip}$ \emph{vs} $\tilde{A}_\mathrm{dip}$ for simulations with a wide range of the geometric parameters $\alpha$ and $K_n$. Two things are notable: for each scattering model, it appears that only a bounded subregion of the $\tilde{\tau}_\mathrm{dip}$-$\tilde{A}_\mathrm{dip}$ plane is accessible by changing the scattering geometry. Moreover, the shape and depth of this accessible domain are different for the different scattering models---differences are even distinguishable between the two palisade models with specular palisade scattering able to access deeper $\tilde{A}_\mathrm{dip}$. These suggest that there are fundamental limits to the strength of AC that can be obtained from backscattering and that these limits are different for the different scattering modes.

If we extend the scatter plots in Figs.~\ref{fig:fig9}(e--h) to three dimensions by also plotting the dimensionless thermal conductivity reduction $1-\tilde{\kappa}_{\infty}$, as is done in Figs.~\ref{fig:fig9}(i--l), we find the surprising result that all of the data points for a particular scattering model lie on a single manifold. The manifold for each scattering model is unique, and are fit well with a simple power law expression relating the thermal conductivity reduction to the $\tilde{\tau}_{\mathrm{dip}}$ and $\tilde{A}_{\mathrm{dip}}$
\begin{equation}
    \label{eq:MC-poower-law-fit}
    1-\tilde{\kappa}_\infty = a + b \tilde{\tau}_\mathrm{dip}^\gamma \tilde{A}_\mathrm{dip}^\eta.
\end{equation}
When fitting this equation to the data for the four scattering models we find values for the fitting exponents of $\gamma= 0.63$, $\eta = 0.22$ and $\gamma= 0.68$, $\eta = 0.23$ for the specular and diffuse palisade models and $\gamma= 0.94$, $\eta = 0.19$ and $\gamma = 0.62$, $\eta = 0.22$ for the specular and diffuse wall models, respectively. 

We have no physical justification for our choice of fitting function, and there are probably other better choices of functional form. However, these observations reveal two important insights. Firstly, there must be some hidden law that relates $\tilde{\tau}_{\mathrm{dip}}$ and $\tilde{A}_{\mathrm{dip}}$ to $\tilde{\kappa}_{\infty}$ --- or rather, if one knows the scattering mode, it suffices to know $\tilde{\tau}_{\mathrm{dip}}$ and $\tilde{A}_{\mathrm{dip}}$ to predict $\tilde{\kappa}_{\infty}$. Second, the unique $\widetilde{\tau}_{\mathrm{dip}}$, $\tilde{A}_{\mathrm{dip}}$, $\tilde{\kappa}_{\infty}$ manifold for each scattering mode means that the scattering modes leave a unique signature in the HCACF. Whilst this is aesthetically pleasing and elegant, it is not terribly useful for any practical purpose in engineering thermal transport properties --- it might, however, offer a means of identifying the nature of phonon back-scattering in different materials systems.

{\subsubsection{Comparison of MD and MC simulations\label{MD-MC}}}

{To make a direct comparison between the MD and MC models we have extended our MC method to explicitly resolve the cylindrical pores in a 3D periodic simulation domain. In this model when a phonon strikes a pore it is back scattered either specularly or diffusely based on the normal vector of the pore surface at the point of impact. However, we can also model the effect of thermalization of phonons when they are scattered at a boundary. In this process, when a phonon collides with a boundary, rather than being immediately backscattered, it is assumed the phonon is absorbed into the local phonon bath and the energy is used for the emission of new phonons at some indeterminate time later. In non-equilibrium MC simulations of such a situation, to obey detailed balance one must keep track of the local temperature rise from the excess of phonons absorbed at the boundaries and use this to appropriately increase the rate of phonon emission from the boundaries. However, in our equilibrium model we can obey detailed balance and model thermalization trivially by following the phonon paths as if they are back scattered (which obeys detailed balance), and then computing the HCACF for each leg of a phonon's trajectory independently (to treat them as uncorrelated).
\begin{figure}[t]
    \centering
    \includegraphics[width=1\textwidth]
    {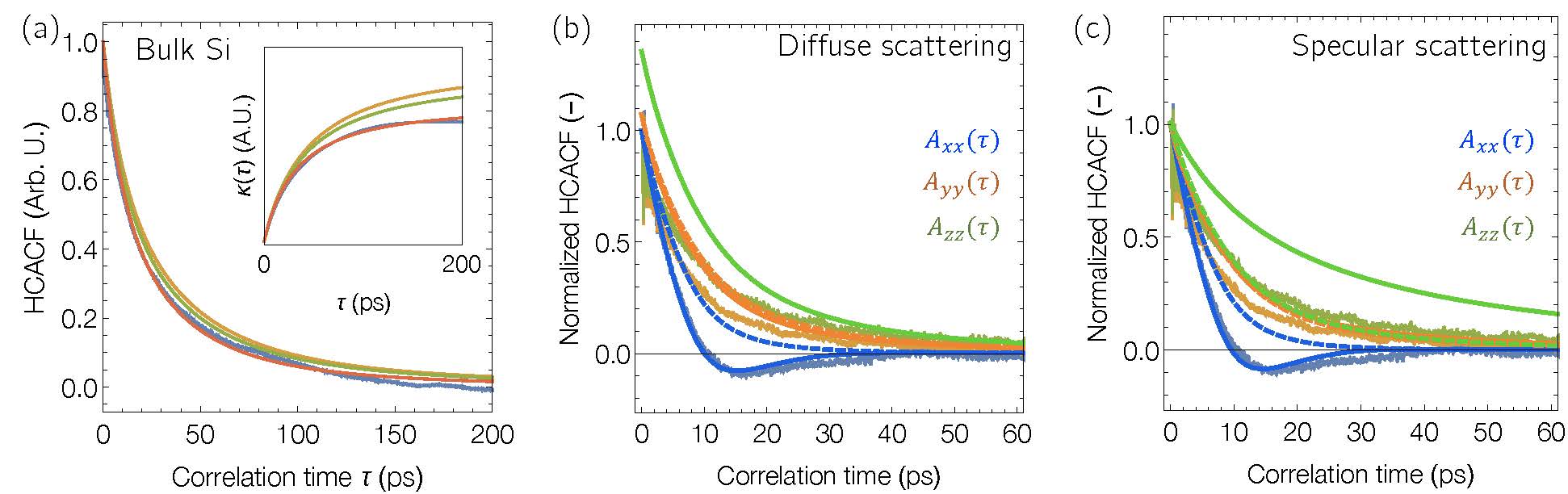}
    \caption{{(a) Comparison of the normalized MD HCACF for bulk Si (blue) and the direct calculation using the phonon dispersion and lifetimes computed from AlmaBTE. The gold and green lines compute $A_\mathrm{bulk}\left(\tau\right)$ assuming, quantum mechanical and classical occupancy, respectively. The red line is classical occupancy and lifetimes scaled down by 0.9 and truncated to 700 ps. The inset shows the integrated HCACF. Plots (b) and (c) show the HCACF in all three directions (normalized by $A_{xx}\left(0\right)$) for nanoporous Si with the geometry in Fig. 1(c). Plot (b) shows the comparison with the MC model with diffuse scattering at the pores, while (c) is for specular scattering at the pores. In all cases the lines plotted in blue, orange and green are for the $x$, $y$ and $z$ directions respectively. The noisy lines are the MD results. The smooth solid lines are $A_{NP}\left(\tau\right)$ from MC simulations with all phonons backscattered at pores. The smooth dashed lines are from MC simulations in which the phonons thermalize when they scatter at pores. Without backscattering there is no super-suppression of the MFP.}}
    \label{fig:MD-MC}
\end{figure}}

{The MD simulations account for the full (classical) phonon spectrum of the Si and the resulting HCACF is the superposition of contributions from all modes. We can compute the full spectrum HCACF for bulk Si, $A_\mathrm{bulk}\left(\tau\right)$, directly from the analytic model as the sum over phonon wave vectors, $q$, and polarizations, $p$, in the Brillouin zone:
\begin{equation}
    A_\mathrm{bulk}\left(\tau\right)=\frac{k_B T^2}{V} \sum_p \sum_q \frac{1}{3} C_{qp}  v_{pq}^2 e^{-\frac{\tau}{\bar{\theta}_qp}},
\end{equation}
where the volumetric specific heat, $C_{qp}$, group velocity, $v_{qp}$, and mean lifetime, $\theta_{qp}$, of the phonon modes are computed from the 2nd and 3rd order stiffness constants for the Si modeled with the Tersoff potential using the AlmaBTE~\cite{carrete2017almabte} to compute the phonon-phonon scattering matrix elements. The normalized $A_\mathrm{bulk}\left(\tau\right)$ computed in this way is plotted in Fig. \ref{fig:MD-MC}(a) along with the HCACF for bulk Si obtained from MD. In the MD simulations the modes are occupied classically and it can be seen that $A_\mathrm{bulk}\left(\tau\right)$ computed using classical values of $C_{qp}$ better matches the MD data than the quantum mechanically occupied $C_{qp}$. The match with MD is not perfect, AlmaBTE predicates there to be a significant number of long lived phonons, particularly along high symmetry directions for which the volume of $q$-space in which scattering selection rules are satisfied is small. These yield a long tail that is not present in the MD simulation. A good match between the MD HCACF and $A_\mathrm{bulk}\left(\tau\right)$ can be achieved by truncating the longest phonon lifetimes to 700 ps, and by scaling all of the AlmaBTE lifetimes down by a factor of 0.9. This may be rationalized as accounting for the extra phonon-phonon scattering present in MD due its classical mode occupancy.} 

{Superposition of HCACFs from individual phonon modes can also be used with our MC simulations to compute the full spectrum HCACF for the nanoporous Si. For a grey phonon population in nanoporous Si, the only material property that affects the shape of the HCACF is the average length of the phonons’ MFP in comparison to the pore spacing. We can thus perform a series of simulations for a population of unit phonons (phonons with unit velocity and unit specific heat) that systematically compute the HCACF over a sweep of phonon lifetimes, $\theta$, and construct a tabulated master function, $\tilde{A}_{NP}\left (K_N,\tilde{\tau}\right)$. This gives the correlation function for a population of phonons in a system with Knudsen number, $K_N$, at the dimensionless correlation time $\tilde{\tau}=\tau/\theta$. We have computed this over five orders of magnitude in $K_N$. Using this, the full spectrum HCACF for the nanoporous Si, $A_{NP}\left(\tau\right)$, is constructed using
\begin{equation}
    A_{NP}\left(\tau\right)=\frac{k_B T^2}{V}\sum_p\sum_q C_{pq}v_{pq}^2 \tilde{A}\left(\frac{v_{pq}\theta_{pq}}{d},\frac{\tau}{\theta_{pq}}\right).
\end{equation}}
{There are two assumptions built into this approach: The first is that the Si’s phonon spectrum is forced to be spherically symmetrical. The second is that we have assumed that the scattering at pores is elastic, meaning that the pores introduce no additional coupling between parts of the phonon spectrum. In the language of BTE phonon simulations the system is \say{multi-grey}. The comparison of the HCACF computed in this way with MD is shown in Fig.\ref{fig:MD-MC}(b\&c) for diffuse and specularly scattering pores respectively. In each of these plots, $A_{NP}\left(\tau\right)$ is computed from MC for both the case where all phonons are back scattered from pores, and the case where phonons thermalize when they are scattered. It can be seen that when phonons thermalize at interfaces there is no anticorrelation in the HCACF, demonstrating the central role that backscattering plays in super-suppression of the phonon MFP.} 

{Comparing the Fig.~\ref{fig:MD-MC}(b) and (c) there is little difference between the effect of diffuse and specular scattering on the correlation function along the long direction, $A_{xx}\left(\tau\right)$, although the case with diffuse scattering matched the MD result more closely. However, there are significant differences on the HCACF in the $z$ direction along the axis of the pores. With specular scattering pores should provide no change in the heat current fluctuations in this direction. Comparison to the MD results for the HCACF along $z$ indicate that in the MD simulations scattering is some combination of diffuse and specular scattering.}

\section{Conclusions}

In a previous recent work~\cite{PhysRevB.102.205405} we showed that specific combinations of closely packed pores lead to anticorrelated fluctuations in the heat flux --- termed heat-current anticorrelation effect --- which we attributed to the specular back-scattering of phonons at the pores. It was also observed that this effect can result in a reduction in the thermal conductivity of up to $\sim$80\% based on this close-packing design, compared to equivalent porosity geometries. Wavepacket simulations had previously shown the presence of seemingly specular scattering at the pores, but our recent work shows that negatively correlated heat-flux fluctuations can result from largely diffuse scattering. Moreover, the wavepacket simulations were performed at low temperatures ($\sim$5 K) calling into question the transferability of the results to room temperature. Using a Monte Carlo ray tracing model, we determined that different forms of correlated scattering imprint a unique signature in the autocorrelation function of the heat-flux. This tool can be used in two crucial ways: to qualify the nature of the correlated behavior, as we were wont to do for the close-packed geometries, and to reverse engineer desired scattering behavior based on these unique signatures. Based on these newly modeled conditions, we can infer that the circular pore geometries most strongly resemble a diffusively scattering wall contravening our earlier assumption that specular scattering was the main source of phonon backscattering. In the work presented herein, we further establish that the anticorrelated heat-flux results in the super-suppression or filtering of large mean-free-paths (in the acoustic region of the spectrum) beyond the characteristic length determined by the nanostructured features. This finding provides a clear explanation of what happens because of the negative heat current correlations and offers additional insight on how to engineer and manipulate thermal transport at the nanoscale. In this work, we improve upon a mathematical model for predicting thermal conductivity based on Matthiessen’s rule developed in Ref.~\cite{PhysRevB.100.035409}. The model relies on a parameter, $\mathrm{C}$, which quantifies the scattering strength of a given pore. We showed here that $\mathrm{C}$ can be predicted as a function of the fraction of the region available for phonon propagation (i.e., neck) and the scattering behavior (i.e., backscattering is either absent or present) for both typical scattering and super-suppression conditions. Using the \emph{typical scattering} model to predict thermal conductivities indicated an excellent agreement for geometries where AC is not present, and a systematic thermal conductivity overprediction for geometries exhibiting AC, consistent with the super-suppression of phonon MFPs beyond surface scattering.

\section{Acknowledgments}

MD calculations were performed with the Advanced Research Computing Center Teton Computing Environment (2018) at the University of Wyoming (https://doi.org/10.15786/M2FY47). NN acknowledges funding from the the European Research Council (ERC) under the European Union's Horizon 2020 Research and Innovation Programme (Grant Agreement No. 678763).

\section{Data Availability}

The raw data from these calculations and the codes and scripts for performing them are available upon request from the corresponding authors.

\section{Conflict of Interest}

The authors declare no conflict of interest

\appendix

\section{Appendix}

\subsection{Derivation of Cumulative Thermal Conductivity in Bulk Crystal} \label{sec:apndx-theory-bulk-conductivity}

In this appendix we lay out the detailed derivation of Eq.~\eqref{eq:kappa-bulk-stochastic} in Sec.~\ref{sec:theory-bulk-conductivity}. We start by considering the contribution to the heat current from the the single phonon mode with wave vector $\mathbf{k}$ and polarization $p$
\begin{equation}\label{eq:J_kp}
    \mathbf{J}_{\mathbf{k}p}(t) = \left( \frac{1}{2}+n_{\mathbf{k}p}(t)\right) \frac{\hbar \omega_{\mathbf{k}p} v_{\mathbf{k}p}}{V}.
\end{equation}
Here $n_{\mathbf{k}p}(t)$ is the mode’s occupancy at time $t$. This can be rewritten in terms of the average flux and the instantaneous excursion from the average
\begin{equation}
    \mathbf{J}_{\mathbf{k}p}(t) = \left\langle \mathbf{J}_{\mathbf{k}p} \right\rangle + \left( n_{\mathbf{k}p}(t) -   \left\langle n_{\mathbf{k}p} \right\rangle  \right) \frac{\hbar \omega_{\mathbf{k}p} v_{\mathbf{k}p}}{V},
\label{eq:J(t)_kp}
\end{equation}
where $\left\langle n_{\mathbf{k}p} \right\rangle$ is the Bose-Einstein occupancy
\begin{equation}
    \left\langle n_{\mathbf{k}p} \right\rangle = \frac{1}{e^{\widetilde{\omega}_{\mathbf{k}p}}-1},
\end{equation}
and $\widetilde{\omega}_{\mathbf{k}p}$ is the dimensionless mode frequency $\widetilde{\omega}_{\mathbf{k}p} = \frac{\hbar \omega_{\mathbf{k}p}}{k_B T}$. 

To find the total flux we use Eq.~\eqref{eq:J(t)_kp} and sum over all modes ($\mathbf{k}$ and $p$). In this process, the mean flux from pairs of modes with opposite $\mathbf{k}$ cancel, leaving the net flux to depend only on the sum of occupancy \emph{excursions} from the mean and yielding Eq.~\eqref{eq:J(t)-stochastic} in the main text. If the occupancy fluctuations in one mode are uncorrelated with the fluctuations in the other modes, then when expanding the product of the sum of modal fluxes, the cross-correlations between modes will be zero simplifying the total correlation function to the sum of autocorrelation functions for each mode individually

\begin{equation}
    \left\langle \mathbf{J}(t) \otimes  \mathbf{J}(t+\tau)\right\rangle = \sum_{\mathbf{k'}p}\sum_{\mathbf{k}p}\left\langle \mathbf{J}_{\mathbf{k'}p'}(t) \otimes  \mathbf{J}_{\mathbf{k}p}(t)\right\rangle = \sum_{\mathbf{k}p}\left\langle \mathbf{J}_{\mathbf{k}p}(t) \otimes  \mathbf{J}_{\mathbf{k}p}(t+\tau)\right\rangle
\end{equation}
This, in turn, depends only on the autocorrelation of the occupancy fluctuations 
\begin{equation}
    \left\langle \mathbf{J}_{\mathbf{k}p}(t) \otimes  \mathbf{J}_{\mathbf{k}p}(t+\tau)\right\rangle = \left ( \frac{\hbar \omega_{\mathbf{k}p}}{V} \right )^2 \mathbf{v}^2_{\mathbf{k}p} \left\langle  \left( n_{\mathbf{k}p}(t) -   \left\langle n_{\mathbf{k}p} \right\rangle  \right) \left( n_{\mathbf{k}p}(t+\tau) -   \left\langle n_{\mathbf{k}p} \right\rangle  \right)    \right\rangle,
\end{equation}
where the shorthand notation $\mathbf{v}^2_{\mathbf{k}p} = \left( \mathbf{v}_{\mathbf{k}p} \otimes  \mathbf{v}_{\mathbf{k}p} \right)$ has been used for the tensor product of the group velocity. 

The occupancy $n(t)$ of a phonon mode will be a random stepped function in time as shown in Fig.~\ref{fig:stochastic-model-bulk}(a) with phonon mode holding a constant excitation for some duration before anharmonic interactions with other phonon modes lead to scattering and a reset of the mode’s excitation. This function can be expressed as a sum of boxcar functions that represent the occupancy during the interval between successive scattering events 
\begin{equation}
    n(t) = \sum_{i=1}^{\infty} n_i \Pi_{t_i, t_i+\theta_i}(t),
\end{equation}
where $\Pi_{a,b} (t)$ is the boxcar function (plotted in blue in the top pane of Fig.~\ref{fig:stochastic-model-bulk}(b)
\begin{equation}
\Pi_{a,b}(t) = \begin{cases}
1 & \text{ for } a \leq t< b \\ 
0 & \text{ otherwise }
\end{cases}
\end{equation}
and $n_i$  and $\theta_i$ are the size and duration of the $\mathrm{i^{th}}$ occupancy fluctuation, and the fluctuations abut one another so that  $\theta_i = t_{i+1} - t_i$. 

If the probability of the $\mathrm{i^{th}}$ occupancy $n_i$  and its duration $\theta_i$ are independent from the fluctuations that proceeded it (as in the figure above), then the occupancy is only correlated during the intervals between scattering and so the occupancy correlation function reduces to simply the average of the correlation functions for each excursion with itself
\begin{equation}\label{eq:n}
    \left\langle  \left( n(t) -  \left\langle n \right\rangle   \right) \left( n(t+\tau) -  \left\langle n \right\rangle   \right)  \right\rangle = \sum_{n=0}^{\infty} \int_{0}^{\infty} d \theta' P_{\overline{\theta}}(\theta') R_n(n- \langle n \rangle)^2 A(\tau, \theta'),
\end{equation}
where the subscripts have been dropped temporarily for clarity. Here, $R_n$ is the average rate of scattering events that leave the mode with occupancy $n$, and $P_{\overline{\theta}}(\theta)$ is the probability that a fluctuation of occupancy $n$ survives for time $\theta$ before scattering given that the average duration of fluctuations is $\overline{\theta}$. The term $A(\tau,\theta)$ is the autocorrelation function of a single boxcar function
\begin{equation}
    A(\tau, \theta) = \int_{0}^{\infty} dt' \Pi_{a, a+\theta}(t') \Pi_{a, a+\theta}(t'+\tau) = \theta \left( 1- \frac{\tau}{\theta}\right) H(\theta-\tau),
\end{equation}
with $H(x)$ is the Heaviside theta function. The integral of the boxcar’s autocorrelation function is 
\begin{equation}
    C(\tau, \theta) = \int_{0}^{\infty} d\tau' A(\tau', \theta) = \left( \theta \tau - \frac{\tau^2}{2}\right) H(\theta-\tau)+\theta^2(1-H(\theta-\tau)).
\end{equation}
The upper pane of Fig.~\ref{fig:stochastic-model-bulk}(b) shows the boxcar function for a single fluctuation, with the lower pane showing its autocorrelation function and integral. The integrated ACF converges to
\begin{equation} \label{eq:C}
    C(\infty, \theta) = \frac{1}{2} \theta^2.
\end{equation}
If the scattering processes that lead to the occupation fluctuations are random, then occupation times $\theta$ are drawn from the Poisson distribution of waiting times
\begin{equation}\label{eq:P}
    P_{\overline{\theta}}(\theta) = \frac{1}{\overline{\theta}} e^{-\theta/\overline{\theta}}.
\end{equation}
The rate, $R_n$, of scattering into occupancy $n$ is related to the thermodynamic probability $P_n$ of finding the mode in its $\mathrm{n^{th}}$ state of occupancy by
\begin{equation}\label{eq:R}
    R_n = \frac{P_n}{\overline{\theta}},
\end{equation}
where $P_n$ is the probability distribution for the canonical ensemble
\begin{equation}
    P_n = e^{-n\widetilde{\omega}} \left(1 - e^{-\widetilde{\omega}} \right).
    \label{eq:Pn}
\end{equation}
Using Eq.~\eqref{eq:P}--\eqref{eq:Pn} in Eq.~\eqref{eq:n} and performing the integral over lifetimes and summing over $n$ gives

\noindent Performing the averaging over all possible occupancies gives
\begin{equation}
    \left\langle  \left( n(t) -  \left\langle n \right\rangle   \right) \left( n(t+\tau) -  \left\langle n \right\rangle   \right)  \right\rangle = \frac{1}{\overline{\theta}} \frac{e^{\widetilde{\omega}}}{(e^{\widetilde{\omega}}-1)^2}\int_{0}^{\infty} d\theta'A(\tau,\theta)e^{-\theta'/\overline{\theta}},
\end{equation}
and averaging over all possible fluctuation durations gives 
\begin{equation} \label{eq:ave}
\left\langle  \left( n(t) -  \left\langle n \right\rangle   \right) \left( n(t+\tau) -  \left\langle n \right\rangle   \right)  \right\rangle = \frac{e^{\widetilde{\omega}}}{(e^{\widetilde{\omega}}-1)^2} e^{-\theta'/\overline{\theta}}
\end{equation}
The cumulative thermal conductivity tensor is then
\begin{equation}\label{eq:kappa_tau}
\mathbf{\kappa}(\tau)= \frac{V}{k_B T^2} \sum_{\mathbf{k}p} \left( \frac{\hbar \omega_{\mathbf{k}p}}{V} \right)^2 \frac{e^{\widetilde{\omega}}}{(e^{\widetilde{\omega}}-1)^2} \mathbf{v}^2_{\mathbf{k}p} \int_{0}^{\tau} d\tau'e^{-\tau'/\overline{\theta}_{\mathbf{k}p}},
\end{equation}
which after some mathematical manipulation simplifies to Eq.~\eqref{eq:kappa-bulk-stochastic} in the manuscript. 

\subsection{Heat Flux Decomposition} 
\label{sec:Heat-Flux-Decomposition}

\begin{figure*}[t]
    \centering
    \includegraphics[width=1\textwidth]{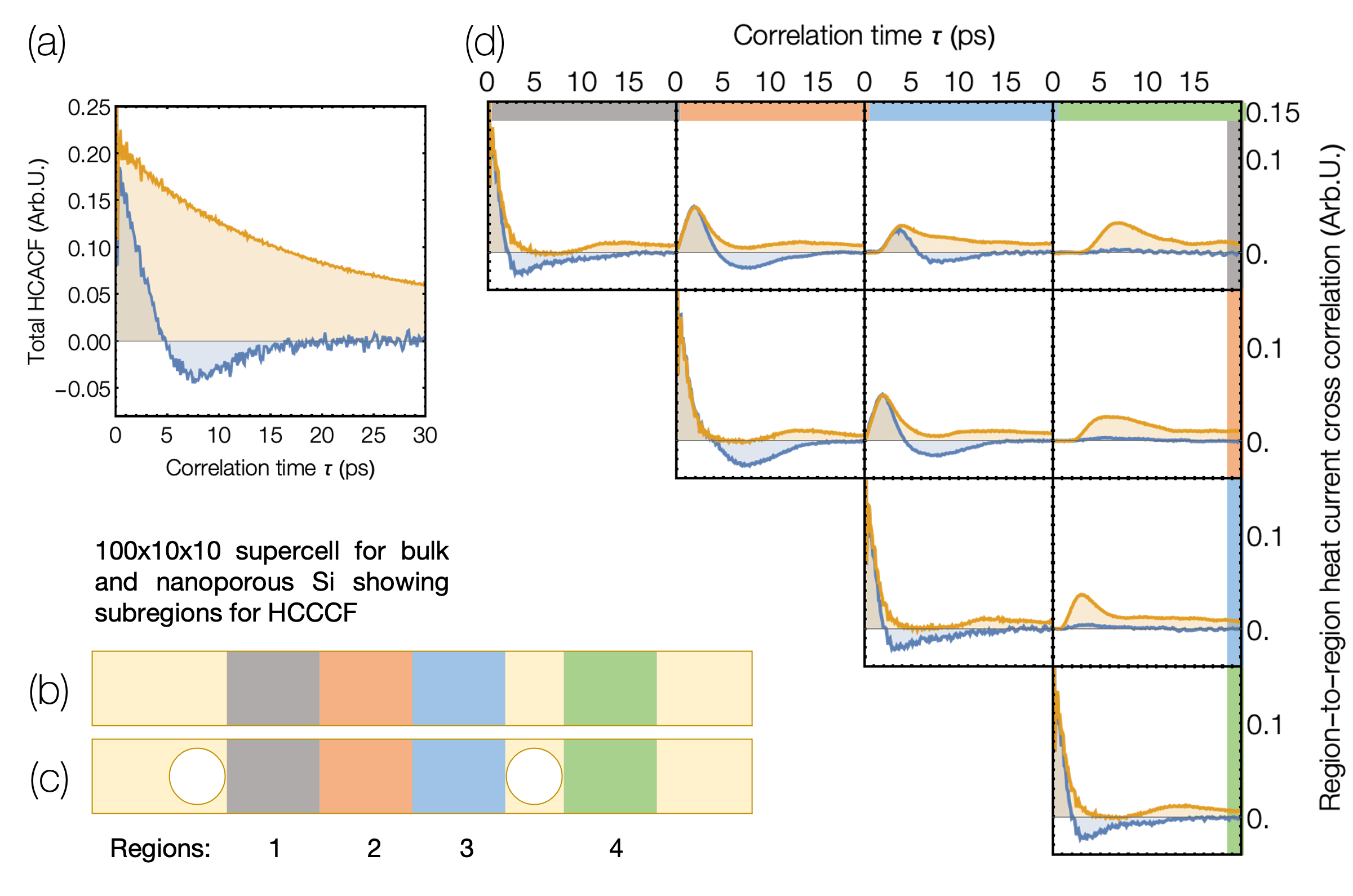}
    \caption{Decomposition of the HCACF. Panel (a) shows the Green-Kubo HCACF for the bulk (gold) and nanoporous (blue) systems. Panels (b) and (c) show the regions used for the spatial decomposition of the heat current in MD simulations of bulk and nanoporous Si, respectively. The matrix of plots in (d) shows the cross-correlations of the heat flux in pairs of regions for the bulk (gold) and nanoporous (blue) systems. The color-coding of the rows and columns matches that of the regions in (b) and (c). The diagonal elements correspond to the regional autocorrelation functions and the remaining figures to the cross-correlation functions. The matrix is symmetric and so the four of the plots in the lower half have been omitted and the space used to show the total HCACF for the two systems.}
    \label{fig:cylinder-CCFs}
\end{figure*}

If the AC affect is due backscattering of phonons at pores we would expect countervailing heat current fluctuations to be spatially, as well as temporally, correlated. Unfortunately, the  HCACF obtained from MD simulations gives no information on the spatial relationship between thermal fluctuations. To overcome this, we have performed simulations in which we decompose the total instantaneous heat flux into contributions from separate sub-regions in the simulation, and used these to compute the  auto- and cross-correlation of the heat current in different regions. These simulations were performed for Si with and without cylindrical pores and the comparison of the results is shown in Fig.~\ref{fig:cylinder-CCFs}. The calculations, which are described in detail in SI Sec.~\ref{sec:computational-methods}, require a few additional steps compared to the calculation of the total HCACF because, unlike the total center of mass, the center of mass of each sub region is not fixed.
    
The regions used for the spatial decomposition in Fig.~\ref{fig:cylinder-CCFs}(c) were chosen carefully, with regions 1--3 located between the two pores in the supercell, and region 4 located to the right of the second pore in the supercell. In the porous Si, regions 1, 3 and 4 are geometrically equivalent. The same regions were used in the simulations of pristine Si, Fig.~\ref{fig:cylinder-CCFs}(b), but in this case all regions are geometrically identical. The matrix of auto- and cross-correlation functions is shown in  Fig.~\ref{fig:cylinder-CCFs}(d), with the region's indicated using the same color coding in the side bar as in Fig.~\ref{fig:cylinder-CCFs}(b\&c) with the 

In bulk Si, all correlation functions between a region with itself (the plots on the diagonal of the matrix) shows two distinct peaks, one at short correlation times corresponding to a heat flux fluctuation with itself, and a much broader and less pronounced peak at a longer correlation interval that comes from phonons that travel through the region and continue around the periodic boundary and back to the region a second time. The heat current cross-correlation functions (HCCCF) between pairs of regions show a time delay before correlation occurs that corresponds to the time of flight for phonons to travel between the regions. For regions that are further away the first peak in correlation becomes broader, reflecting the distribution of $x$ projected velocities of the heat-carrying phonons. 

The HCACF and HCCCF for the Si with cylindrical pores are markedly different from those of the bulk Si. In regions 1, 3 and 4 about the pores, the HCACF decays rapidly to a sharp region of anticorrelation which is dominated by around half of the phonons being backscattered off the immediately adjacent pore. In principle, there could be a second later peak and dip that occurs later due to phonons that are reflected from the more distant pore, but the velocity dispersion or changes in direction of the scattered phonons mean that this is not visible or not present. Region 2 is located right between the pores. It too shows a negative region in the HCACF, but at a later interval, due to the three times longer distance of travel to the pores. That said, the anticorrelation is stronger than for region 1 as it is contributed to by both the left and right traveling phonons. Again, it is notable that there is no secondary positive correlation from phonons that are scattered twice (once at the left and right pores) and then pass through region 2 for the third time. This would be the equivalent to a secondary HCACF dip in the HCACFs in bulk Si. This is significant as it means that phonons are uncorrelated after their second backscattering event.   

The cross-correlations of the nanoporous systems between regions 2 and 1 match the bulk behavior initially, but the initial positive correlation is followed by a long interval of anticorrelation. Importantly, in the nanoporous system, regions 1 and 2 of the HCCCF do not have a secondary peak as seen in the bulk Si which is due to phonons that travel around the periodic boundaries of the system. In the nanoporous Si, the route from region 1 to 2 wrapping around the boundaries requires passing two ranks of pores. In fact, the obstacle that the pores present for phonon scattering is seen most clearly in the HCCCFs between region 4 and regions 1, 2, and 3. In these plots, there is no discernible correlation, either positive or negative.

Together these cross-correlations show definitively that (1) backscattering occurs at pores and not at some other region of the system, (2) that a single row of pores [at least of the size in Fig.~\ref{fig:cylinder-CCFs}(c)] is sufficient to prevent the propagation of heat carriers and (3) that the secondary scattering of phonons is uncorrelated. This resolves an important question from our earlier work where we had simulated the scattering of individual phonon wavepackets colliding with pores~\cite{PhysRevB.102.205405}. The wavepacket simulations revealed that the transmission of phonons past the pores is wavelength dependent, but that when reflection occurs wavepackets scatter back and forth multiple times. This seemed contradictory to the single dip seen in the HCACF. However, the non-equilibrium wavepacket simulations were performed at ultralow temperature, and so lacked a full phonon bath. The decomposed equilibrium MD simulations in Fig.~\ref{fig:cylinder-CCFs} complete the picture, showing that at 300 K the heat-carrying phonons are intrinsically sufficiently long-lived to undergo multiple scattering events, but that they lose (temporal) coherence after a secondary scattering.

\bibliographystyle{unsrt}  
\bibliography{references} 

\beginsupplement
\newpage
\paragraph{SUPPORTING INFORMATION}

\section{Computational Methods} \label{sec:computational-methods}

Our study focuses on two materials systems: silicon and a pseudo-material with a grey phonon population that we model in our ray tracing simulations. The latter system enables us to examine the effect of correlated scattering in isolation and unobfuscated by the presence of a broad phonon spectrum. Below we describe the details of the different modeling approaches.

\paragraph{Green--Kubo Calculations}

The phonon-mediated heat current correlation is related to the lattice thermal conductivity through the Green-Kubo (GK) formalism. The GK equation is the reduced form of the fluctuation-dissipation theorem in the linear-response region and relies on the assumption that the same mechanisms or processes, by which a system responds to a stimulus or perturbation (e.g., temperature gradient) are responsible for the system's response to local fluctuations (e.g., instantaneous heat flux) in equilibrium. Mathematically, this means the thermal conductivity can be calculated from equilibrium MD simulations using Eq.~\eqref{eq:kappa_GK} if one computes $\left\langle \mathbf{J}(t) \otimes  \mathbf{J}(t+\tau)\right \rangle$. Due to the increase in the error of the heat current ACF over time, it is customary to truncate the ACF. In this study, the ACF is truncated at 150 ps for all systems, except the pristine geometry for which the cutoff was set to 500 ps (due to a much slower relaxation process). A more thorough discussion on the topic of error mitigation in the ACF can be found in Ref.~\cite{de2017method}, and a discussion/ justification on cutoff selection, can be found in Ref.~\cite{de2019large}. Due to the real-time nature of the simulations, no explicit definition of the phonon quasiparticle is needed, and the harmonic and anharmonic interactions of phonons are implicitly captured through the choice of interatomic potential. By not treating phonons explicitly as particles, MD captures wave effects, such as coherence/ decoherence~\cite{PhysRevLett.128.015901}, and merges the phonon nature of waves and particles. We opted to use both the Stillinger–Weber (SW)~\cite{stillinger1985computer} and Tersoff~\cite{tersoff1988empirical} potentials to tease out potential-dependent computational artifacts. The MD simulations were performed with the Large-scale Atomic/ Molecular Massively Parallel Simulator (LAMMPS)~\cite{plimpton1995fast}. {We remark that LAMMPS underestimates the thermal conductivity of materials described by many-body potentials, however the underestimation is most problematic in low-dimensional materials. For bulk materials such as Si, LAMMPS yields reliable prediction of thermal conductivity. We note that, even if there was a problem with the way that LAMMPS computes the thermal flux, this would only change the amplitude of the heat current and would not produce the qualitatively different behavior that we see between heat flow in bulk and porous Si~\cite{PhysRevB.92.094301}}. The results were averaged for 10--20 sets of simulations to mitigate the large uncertainty in the GK approach. Simulation supercells range in sizes, and details about the size of simulation cells are therefore included in the text for each case. Each supercell contains pores that perforate through the supercell [see inset in Fig.~\ref{fig:spectra}(e)]. In all cases, atoms were equilibrated to 300 K. The initial step was to bring the systems to room temperature in the isothermal, isobaric ensemble (NPT), allowing for thermal expansion. This was done over 125 ps. Then, the systems were equilibrated in the microcanonical ensemble (NVE) for an additional 125 ps. The GK calculations were performed over 10 ns, also in NVE. The simulations were performed using a 1-fs interval. All transport properties reported in this work were done along the longest side of each simulation cell, which corresponds to the $\langle 1 0 0 \rangle$ Si crystal direction [see inset in Fig.~\ref{fig:spectra}(e)]. The geometries in Fig.~\ref{fig:fig1} were averaged over 5 sets of simulations, using the Tersoff potential. The supercells are indicated in the figure. To identify the spectral composition of the heat flux, the discrete cosine (Fourier) transform of the HCACF, which is equivalent to the same transformation of the heat flux, was computed for a series of geometries. To help smooth the spectra for clarity, a Gaussian filter with a (pixel) width of 15 was used for the data in Figs.~\ref{fig:spectra}(c), (d), (h), and (i). The Stillinger--Weber potential was used for this set of calculations. The resulting HCACF spectra are plotted as a function of $\tau^{-1}$, where $\tau$ is the relaxation time and, therefore, inversely proportional to the phonon MFP. 

\paragraph{Classical Phonon Wavepacket Model}

A phonon wavepacket analysis was performed to help elucidate the nature of heat transport in the vicinity of the pores. We formed a set of propagating wavepackets by linearly superposing phonon plane waves weighted by Gaussian distributions around given wave vectors. Mathematically, a Gaussian wavepacket is defined as
\begin{equation}\label{eq:WP}
    u_{l j \mu \gamma} =   \sum_q A_o \left( \frac{1}{\sigma \sqrt{2\pi}} \right)^2 e^{\left( \frac{q-q_o}{\sigma \sqrt{2}}\right)^2} \epsilon_{j \mu} e^ {-i\left(r_l q+\omega_\gamma t \right)}.
\end{equation}
Here, $u_{l j\mu \gamma}$ is the displacement of the $j^{th}$ atom in the $l^{th}$ unit cell along a direction $\mu$ (in $x$, $y$ or $z$) for a given mode, $\gamma$. $A_o$ is the amplitude of the wavepacket, which can be tuned to the desired wavepacket energy. The wavepacket is spaced around a given wave vector, $q_o$, with uncertainty in momentum space specified by $\sigma$. The term $\omega_\gamma$ is the frequency of the mode $\gamma$ at  $q_o$, and $\epsilon_{j \mu}$ is the eigenvector of the $j^{th}$ atom along $\mu$ at the selected mode, $\gamma$. $r_l$ is a vector that points to the $l^{th}$ unit lattice, and $t$ is the time. The sum over $q$ is performed for all wave vectors in the first Brillouin zone that are commensurate with the compute cell. The phonon wavepackets are localized around wave vectors $q_o$ along the $\langle 1 0 0 \rangle$ crystal direction for both the longitudinal and transverse acoustic modes $\gamma$. A $800 \times 10 \times 10$ Si supercell with very fine uncertainty in momentum space ($\sigma = 0.01\ \mathrm{nm}^{-1}$) is used. The initial position of the atoms is computed using Eq.~\eqref{eq:WP} and the initial velocity is computed from the time derivative of $u_{l j\mu \gamma}$ ($v_{l j\mu \gamma} = \frac{d}{dt} u_{l j\mu \gamma}$). The system is initially relaxed at 0 K then the wavepackets are added with $A_o$ for each wavepacket tuned so that the temperature of the system is raised by around $\sim$5 K. Performing the wavepacket simulations at low energy ($\sim$5 K) helps keep phonon thermalization at bay, such that the acoustic frequencies don’t easily decay into other modes/frequencies due to anharmonicity. This allows us to observe the scattering behavior of phonons at the surfaces of the nanopores, as they are less likely to be obfuscated by anharmonic effects. 

\begin{figure*}[t]
    \centering
    \includegraphics[width=1\textwidth]{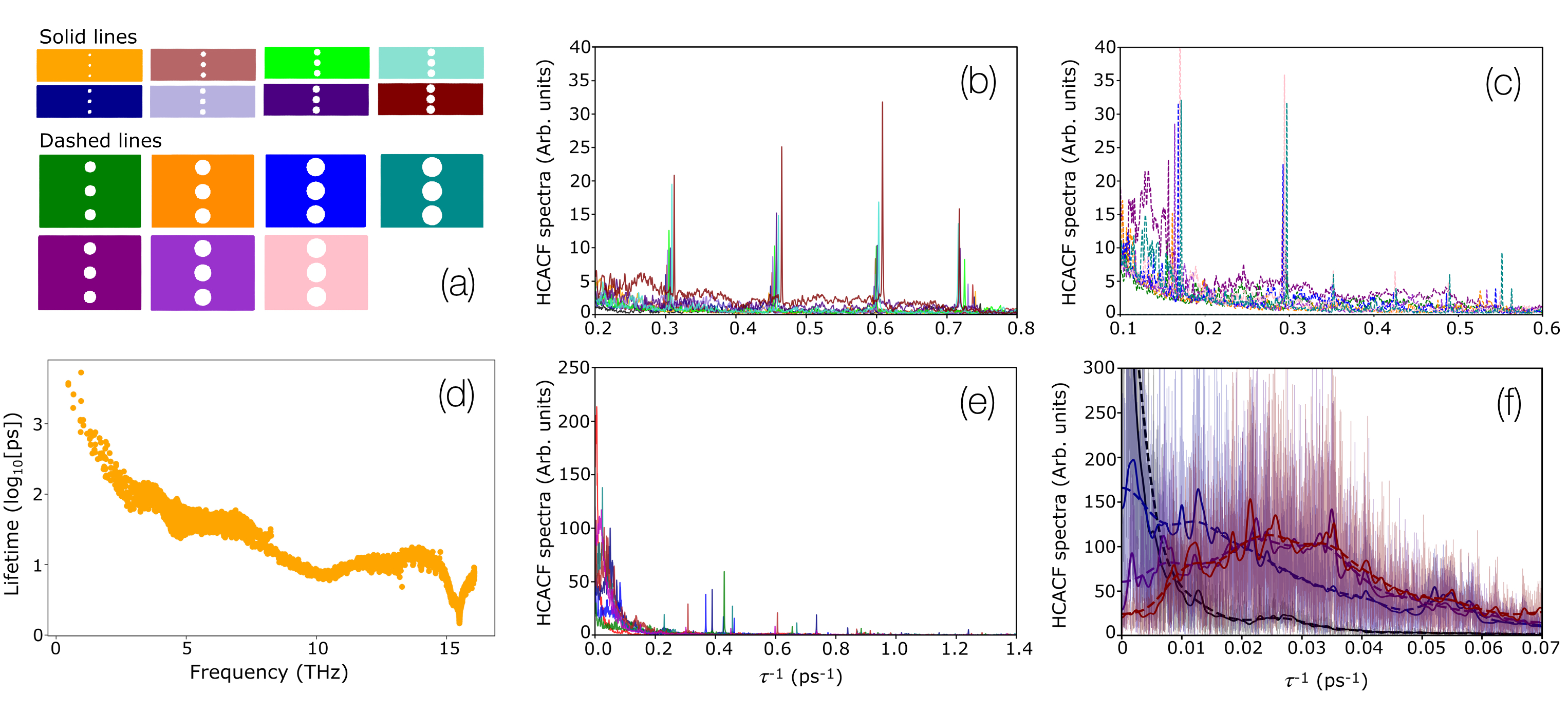}
    \caption{(a) Schematic view of the simulation domains in Fig.~\ref{fig:spectra}. The corresponding HCACF spectra for the geometries with narrow ($\mathrm{100\times10\times10}$ supercell geometries) and wide ($\mathrm{100\times24\times10}$ supercell geometries) width are plotted in panes (b\&c), respectively. Panel (d) shows phonon lifetime, $\tau$, versus frequency. The spectra perpendicular to the transport direction for $\mathrm{100\times10\times10}$ supercell geometries are plotted in pane(e). Panel (f) Original (i.e., not filtered/smoothed) spectra (solid faint purple and red lines) for two of the highest HCACF dip geometries as shown in (a)--(c), including the spectra filtered with a Gaussian of width 15 (solid dark lines), and a much smoother version (dashed dark lines), of width 100, to help visualize the peaking that occurs in the spectrum of geometry with AC. Black lines correspond to the pristine geometry and the blue line to geometry without a noticeable dip.}
    \label{fig:SI-spectra}
\end{figure*}

\paragraph{Phonon Boltzmann Transport Model}

Mode- and space- resolved Boltzmann transport equation is solved to find the steady-state distribution of phonons moving between an array of pores under an imposed temperature gradient. The effective thermal conductivity is defined as the ratio of the heat flux carried by the phonon distribution divided by the imposed temperature gradient. These simulations were performed using the OpenBTE Boltzmann transport solver~\cite{romano2021openbte}. The second- and third- order interatomic force constants for bulk Si were computed with MD using Phonolammps~\cite{abelcarreras2020phono} and Thirdorder~\cite{li2014shengbte}, respectively. The Tersoff potential is used in this set of calculations. The phonon dispersion was computed from the second-order force constants on a $\mathrm{40\times40\times40}$ point Brillouin zone mesh using AlmaBTE~\cite{carrete2017almabte}. The scattering matrices for three-phonon interactions were computed from the third-order force constants also using AlmaBTE which computes the full three-phonon scattering matrix and uses it to solve the linearized Boltzmann transport equation for phonons.

\section{Spectral Analysis of HCACF and Phonon Supper-suppression Analysis\label{SI:spectral-analysis}}

Fig.~\ref{fig:SI-spectra}(a) shows the complete set of geometries studied in Fig.~\ref{fig:spectra} in the main manuscript. Similar to Fig.~\ref{fig:spectra}, solid lines correspond to $\mathrm{100\times10\times10}$ supercell geometries with narrow width and dashed lines correspond to $\mathrm{100\times24\times10}$ supercell geometries with wider width. Panes (b\&c) show the broader region of HCACF spectra for the narrow and wide width geometries, respectively. Figs.~\ref{fig:SI-spectra}(b\&c) show the HCACF spectra at higher frequencies than depicted in the main text, revealing additional peaks that occur in the porous geometries likely due to phonon scattering at the surface of the pores. Panel (d) shows the zoom-out HCACF spectra for the geometries with narrow width perpendicular to the transport direction and varying longitudinal pore distances. A closer look at the overall spectrum [Fig.~\ref{fig:SI-spectra} (b, c and e)] shows a shift in the peaks of $\theta^{-1}$ beyond roughly 0.1-0.2 ps that correlates closely with the neck width to pore diameter ratio. The structures with larger pores and smaller necks have peaks in HCACF shifted into higher frequencies, compared to the geometries with smaller pores and larger necks. We observe that higher frequency peaks are seen even in geometries with small pores/necking ratio. However, there is no such a peak in the HCACF spectra of pristine Si. This observation suggests that these peaks emerge from the modes carrying heat at the surface of the pores and are not related to the anticorrelation in the HCACF.

\begin{figure*}[t]
    \centering
    \includegraphics[width=1\textwidth]{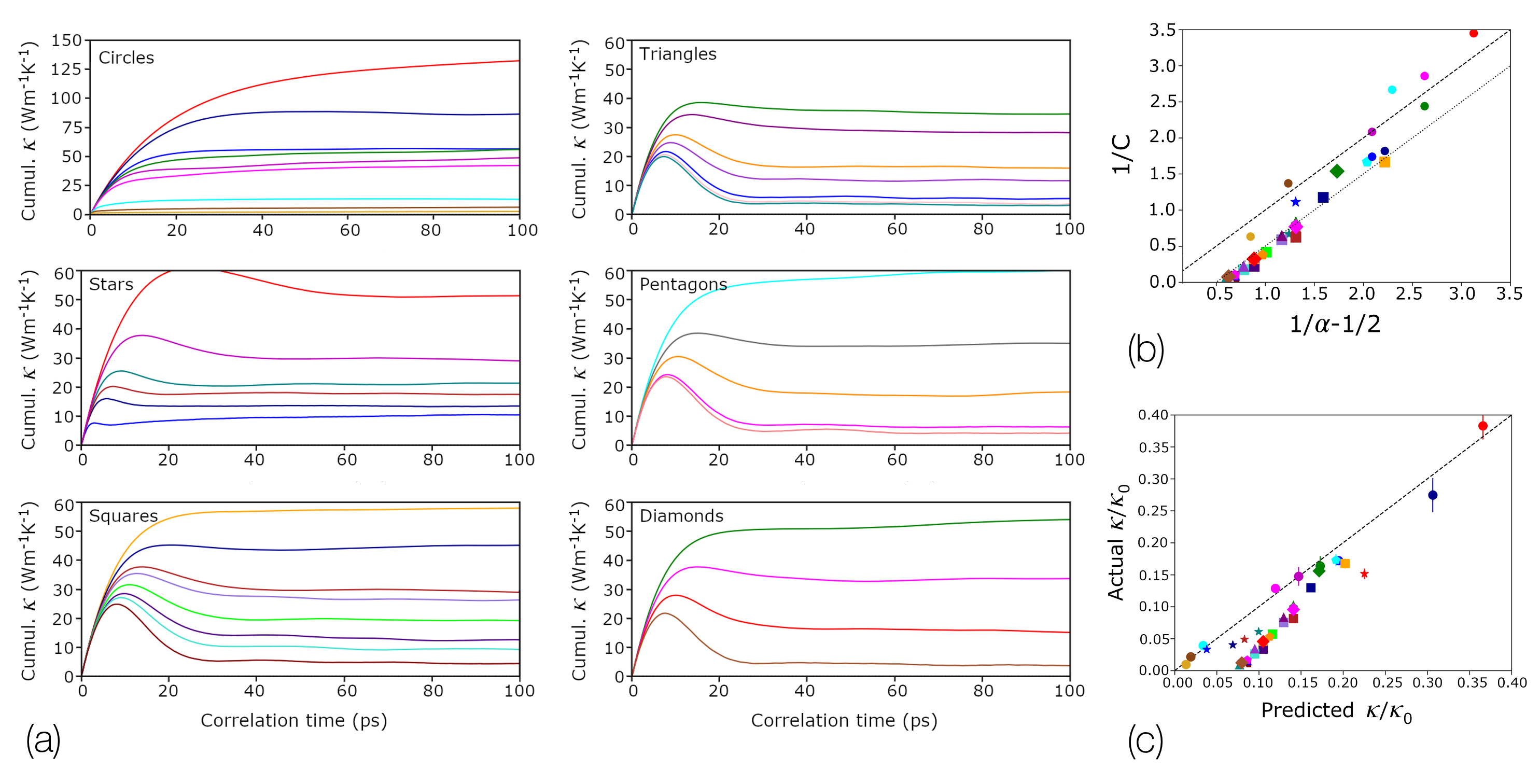}
    \caption{(a) Cumulative thermal conductivity for uniformly distributed pores with varying radius and neck (circles), fixed pore radius (r = 1.5 nm) but varying pore spacing (stars), fixed pore-pore spacing of 5.43 nm perpendicular to the heat current with pore radii vary between 1 and 2.5 nm (squares), fixed pore-pore spacing of 13.03 nm perpendicular to the heat current with pore radii vary between 3.6 and 5.92 nm (triangles), fixed pore-pore spacing of 7.60 nm perpendicular to the heat current with pore radii vary between 1.5 and 3.9 nm (pentagons), fixed pore-pore spacing of 10.86 nm perpendicular to the heat current with pore radii vary between 2.44 and 4.83 (diamonds). (b) The corresponding relationship between the scattering parameter $\mathrm{C}$ and the porosity factor. Each set is marked with the label indicated in the cumulative thermal conductivity. (c) Comparison between the predictive thermal conductivities and the MD-computed thermal conductivities.}
    \label{fig:fig5}
\end{figure*}

\section{Super-Suppression of Phonon MFP in MD Model\label{SI:MD-Super-Suppression}}

Fig.~\ref{fig:fig5} shows the cumulative thermal conductivity for the data points in Fig.~\ref{fig:fig3}. Corresponding scattering strengths, $\mathrm{C}$, and normalized thermal conductivities are shown in panels (d\&h).

\section{Scattering Algorithm in the Palisade Ray Tracing Models\label{SI:palisade-model}}

With the palisade model, we aim to more closely mimic the pore geometry simulated in the MD simulations. We assume that the material contains rows of cylindrical pores, each with radius $r$ and aligned with their axis parallel to the $y$-direction. The pores are arrayed in the $z$-direction with spacing $L$ to form a palisade fence. The geometry of the palisade is parameterized with the parameter $\alpha = \frac{2r}{L}$ which is the scattering probability for phonons that meet the palisade head-on. In general, the probability that phonons are scattered by the pores rather than passing between them is a function of the incidence angle, $\chi$, of the phonon trajectory in the $x$-$z$ plane with the $x$-axis. The transmission probability depends on the gap between cylinders that is visible to the phonons as they travel towards the palisade, given by
\begin{equation}
    T(\chi) = 1 - \frac{\alpha}{|\cos \chi|}.
\end{equation}
\begin{figure*}[t]
    \centering
    \includegraphics[width=1\textwidth]{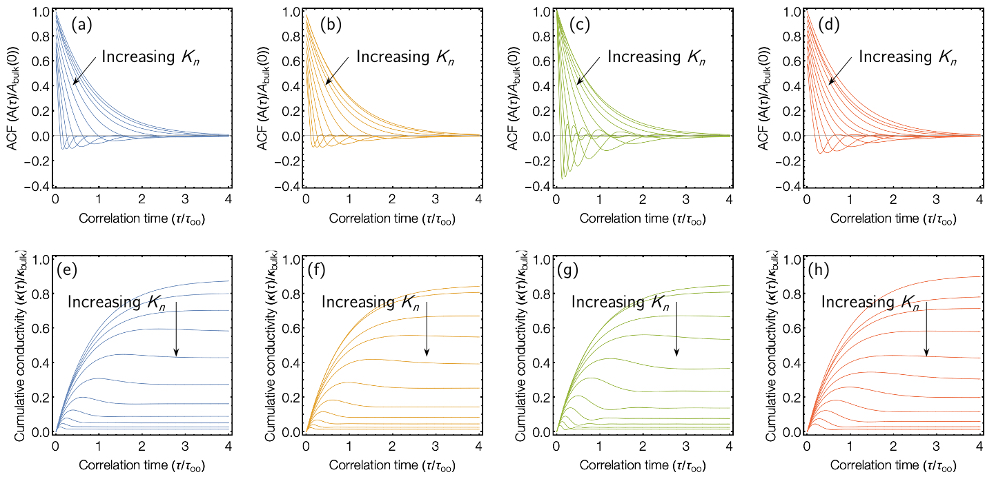}
    \caption{Plots of the HCACF (top row), and corresponding cumulative thermal conductivity (bottom row), for the four scattering models. The specular and diffuse palisade models are plots in blue (a\&e) and gold (b\&f), respectively, while specular and diffuse wall models are plotted in green (c\&g) and red (d\&h), respectively. All plots are for simulations with $\alpha = 0.85$, and the log of the Knudsen number is swept from -1 to 1.  In the top row, the HCACF is plotted normalized by $A_{\mathrm{bulk}}(0)$, the initial value of the HCACF  in the bulk crystal. Similarly, on the bottom row, the cumulative thermal conductivity is normalized by the thermal conductivity of the bulk crystal containing no extrinsic scattering centers.}
    \label{fig:figA2}
\end{figure*}
Here $\tan \chi = \nu_z/\nu_x$. The probability of transmission drops to zero at grazing angles of incidence where the shadow of the cylinder one another. Rather than model the location of cylinders explicitly we consider the probability that an incident phonon strikes a cylinder at a position with normal vector at an angle of $\beta$ relative to the phonon direction. This probability is given by $P(\beta) = \frac{\cos \beta}{\sin \beta_{\mathrm{max}}-\sin \beta_{\mathrm{min}}}$, where $\beta_{\mathrm{min}}$ and $\beta_{\mathrm{max}}$ are the limits to the possible incident angles that a phonon could strike the surface of a cylinder. If there is no shadowing $\beta_{\mathrm{min}} = -\pi/2$, and $\beta_{\mathrm{max}} = \pi/2$. Shadowing occurs at angles when $T(\chi)<0$, and in these cases: 
\begin{equation}
    \beta_{min} = \begin{cases}
 \arcsin \left( 1-\frac{2 \cos \chi}{\alpha} \right ) & \text{ for} \cos \chi \sin \chi > 0\\ 
 -\frac{\pi}{2}& \text{ otherwise }
\end{cases},
\end{equation}

\begin{equation}
    \beta_{max} = \begin{cases}
 -\arcsin \left( 1-\frac{2 \cos \chi}{\alpha} \right ) & \text{ for} \cos \chi \sin \chi < 0\\ 
 \frac{\pi}{2}& \text{ otherwise }
\end{cases}.
\end{equation}
For the specular scattering model, when a phonon’s $x$ position lies on a palisade a pseudorandom number generator is used to decide if the phonon passed through the barrier unscattered or if it struck one of the pillars. If scattering occurs the incidence of a random incidence angle is selected from $P(\beta)$ and the phonon is set on a new trajectory with a new angle
\begin{equation}
    \chi' = \chi+\pi-2\beta.
\end{equation}
Note that this mean-field model of scattering does not resolve the physical size of the pores and the possibility of multiple scattering between neighboring cylinders within a palisade is not considered. The model for diffuse scattering from cylindrical pores uses the same procedure to stochastically select the normal at the point on the surface of the cylinder that the phonon strikes (this has an angle $\chi' = \chi+\pi-\beta$).  A new random direction is then selected for the phonon in the half-space defined by this normal vector. 

\section{Monte Carlo Ray Tracing  Simulations of Correlated Scattering\label{SI:MC-simulations}}

Figure~\ref{fig:figA2} shows the variation in the HCACF predicted from the four scattering models in the ray tracing model as the distance between the ranks of pores is varied. 

\section{Super-Suppression of Phonon MFP in Ray Tracing Models}

\begin{figure*}[h]
    \centering
    \includegraphics[width=1.0\textwidth]{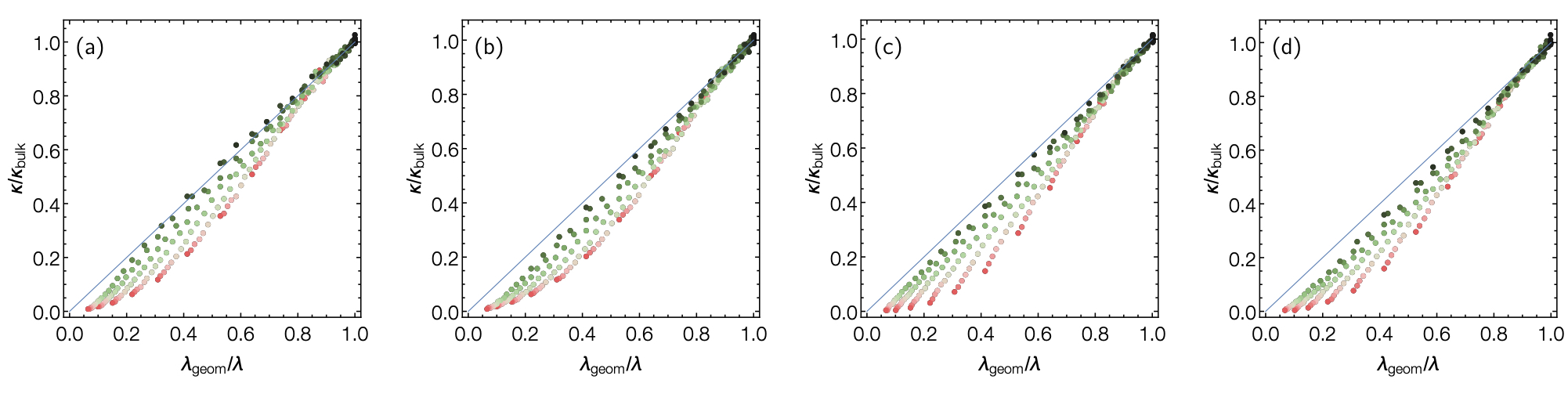}
    \caption{Plots of the thermal conductivity reduction predicted by the ray tracing model versus the geometric reduction in MFP for scattering from (a) a palisade of specularly reflecting cylindrical pores, (b) diffusely scattering cylinders, (b) a specularly reflecting wall, and (d) a diffusely reflecting wall. The data points are colored according to the magnitude of these HCACF dips with red indicating a large dip and green no dip.}    
    \label{fig:SI-super-suppression-MC}
\end{figure*}

Figure~\ref{fig:SI-super-suppression-MC} shows the thermal conductivity reduction $\kappa_p/\kappa_\mathrm{bulk}$ \emph{vs} the geometric mean free distance to an obstacle.

\end{document}